\documentstyle[12pt]{article}

\newcommand{\Od}{{\cal O}}

\begin{document}
\input epsf \renewcommand{\topfraction}{0.8}
\pagestyle{empty} \vspace*{5mm}
\begin{center}
\Large{\bf Brane-skyrmions and wrapped states}
\\ \vspace*{2cm}
\large{ J. A. R. Cembranos, A. Dobado and A. L. Maroto}
\\ \vspace{0.2cm} \normalsize
Departamento de  F\'{\i}sica Te\'orica,\\
 Universidad Complutense de
  Madrid, 28040 Madrid, Spain\\
\vspace*{0.6cm} {\bf ABSTRACT} \\
\end{center} \vspace*{5mm}
 In the context of a brane world and including an induced curvature
term in the brane
action,
we obtain the effective lagrangian for the Goldstone bosons
(branons) associated with the spontaneous breaking of the
translational invariance in the bulk. In addition to the branons,
this effective action has
Skyrmion-like solitonic states which can be understood as holes in the
brane. We study their main properties such as mass and size, the
Skyrmion-branon interaction and their possible fermionic quantization.
We also consider states where
the brane is wrapped around the extra dimensions and their
relation with the brane-skyrmions. Finally, we  extend our results to
higher-dimensional branes, such as those appearing in M-theory,
where brane-skyrmions could also be present.

\noindent
\begin{flushleft} PACS: 11.25Mj, 11.10Lm, 11.15Ex \\
\end{flushleft}
\newpage
\setcounter{page}{1} \pagestyle{plain}
\textheight 20 true cm
\section{Introduction}

In recent years, the old Kaluza-Klein \cite{KK} idea of having
extra dimensions has received a lot of attention. The main new idea
that triggered this revival was the suggestion that our world
could be a brane embedded in large extra dimensions, in such a way
that the standard fields are confined to live in the brane, but
gravitons are free to move on the whole bulk D-dimensional space
\cite{Hamed1}. Probably the most appealing property of this
scenario is that the true D-dimensional gravitational scale $M_D$
could be as small as a few TeV, thus making possible to have
gravitational effects reachable in the next generation of colliders.
In fact many works have been devoted recently to the study of
the phenomenological implications of the brane-world scenario
\cite{Hamed2}. In addition, some of the old problems of the
Standard Model can be reconsidered in a completely new way (see
\cite{Abdel} and references therein). Usually one assumes that the
new $N$ dimensions are
compactified in some N-dimensional manifold $B$ with a typical
size $R_B$. At low energies, the relevant degrees of freedom are
the Standard Model fields on the brane and the gravitons 
 with the corresponding infinite tower of Kaluza-Klein (KK)
partners (see for instance \cite{Balo} for a review of 
Kaluza-Klein theory and models) and finally the brane own
excitations or branons. The interactions of the graviton sector with
the Standard Model fields have been analyzed in several papers
 \cite{Giudice}. However, the presence in the bulk of the brane  
in its ground state typically breaks spontaneously some of the
isometries of the space $B$. The brane configurations obtained 
from the brane ground state by
isometry transformations corresponding to the broken isometries
 (typically translations) produce
equivalent ground states and thus, the parameters describing these
transformations can be considered as the Goldstone bosons (GB)
fields of the isometry breaking (zero-mode branons). In fact when
the brane tension scale $f$ is much smaller than the fundamental
one, the non-zero KK modes decouple from the GB modes \cite{GB}
and then it is possible to make a low-energy effective description
of the GB dynamics \cite{Sundrum}. The phenomenological effects of
these branons have been considered in \cite{Kugo}.

On the other hand, in the standard Kaluza-Klein approach, the
isometries of the space $B$ are understood as gauge symmetries in
the four dimensional space-time. Since the GB are associated to
the spontaneous breaking of some of those symmetries, the Higgs
mechanism must take place. However the gauge boson (graviphoton)
masses produced in this way are very small (typically $M\sim f^2/M_P$),
 so that one can advocate the Equivalence Theorem
\cite{ET} and neglect the graviphoton masses for practical
purposes.

In addition to the branons, the brane can support also a new set of
topological states. These states are defects that appear due to
the non-trivial homotopies of the vacuum manifold \cite{Shifman}.
In particular the authors of this reference have considered the
case of string and monopole defects on the brane, corresponding
to non-trivial first and second homotopy groups.

In this work we are interested in another kind of defects of
topological nature  related to the Skyrme model \cite{Skyrme}. In
this model, the baryons are understood as topological solitons
that appear in the low-energy pion dynamics described by a chiral
lagrangian, the baryon number being identified with the
topological charge. This model has provided a very successful
description of the baryon properties \cite{Witten2}. In a recent
work \cite{Dobado}, two of us derived the effective action for the
brane GB or zero-mode branons starting from a
Dirac-Nambu-Goto-type action for the brane. This effective action
is formally similar to the chiral lagrangians used for the low-energy
description of the chiral dynamics \cite{Weinberg} or even of the
symmetry breaking sector of the Standard Model in the strongly
coupled case \cite{DH} (see \cite{Book} for a review of effective
lagrangians and their applications to the Standard Model and
Gravitation). Therefore it is quite natural to wonder about the
possibility of having chiral solitons (Skyrmions) arising from
this effective action. As we will show the answer is positive. In
the following we will study in detail those brane-skyrmions, their
physical and geometrical interpretation, their main properties and
their relation to wrapped states.

The plan of the paper goes as follows: In Sec.2 we introduce our
set up and extend our previous results for the brane GB effective
action starting from a generalized brane action that includes an
induced scalar curvature term. This term will be essential in
order to determine the brane-skyrmion size. In Sec.3 we use the
effective action to obtain the vacuum equations for the brane. We
also consider the effects of small deviations from the ideal
symmetry breaking pattern which will give rise to small mass terms
for the branons. In Sec.4 we give the equations for the
brane-skyrmions and compute analytically their size and mass in
terms of the different parameters.
In Sec.5 we give more general numerical results. In Sec.6 we
extend our analysis to higher-dimensional brane-skyrmions. This
kind of solutions could have some relevance for pure M-theory
beyond the brane-world scenario. Sec.7 is devoted to the
interactions between the brane-skyrmions and branons and  the
possible fermionic quantization. The effect of the possible branon
masses on the brane skyrmions is considered in Sec.8. In Sec.9 we
consider another  set of brane states (wrapped states) and study
their relation with the brane-skyrmions. Finally, in Sec.10 we set
the main results of our work and the conclusions.

\section{The effective action for the branons}

Let us start by fixing the notation and the main assumptions used
in the work.  We consider that the four-dimensional space-time
$M_4$ is embedded in a $D$-dimensional bulk space that for
simplicity we will assume to be of the form $M_D=M_4\times B$,
where $B$ is a given N-dimensional compact manifold so that
$D=4+N$. The brane lies along $M_4$ and we neglect its
contribution to the bulk gravitational field. The coordinates
parametrizing the points in $M_D$ will be denoted by
$(x^{\mu},y^m)$, where the different indices run as $\mu=0,1,2,3$
and $m=1,2,...,N$. The bulk space $M_D$ is endowed with a metric
tensor that we will denote by $G_{MN}$, with signature
$(+,-,-...-,-)$. For simplicity, we will consider the following
ansatz:
\begin{eqnarray}
 G_{MN}&=&
\left(
\begin{array}{cccc}
\tilde g_{\mu\nu}(x)&0\\ 0&-\tilde g'_{mn}(y)
\end{array}\right).
\end{eqnarray}
In the absence of the 3-brane, this metric possesses an isometry
group that we will assume to be of the form $G(M_D)=G(M_4)\times
G(B)$. The presence of the brane spontaneously breaks this
symmetry down to some subgroup $G(M_4)\times H$. Therefore, we can
introduce the coset space $K=G(M_D)/(G(M_4)\times H) =G(B)/H$,
where $H\subset G(B)$ is a suitable subgroup of $G(B)$.

The position of the brane in the bulk can be parametrized as
$Y^M=(x^\mu, Y^m(x))$, where we have chosen the bulk coordinates
so that the first four are identified with the space-time brane
coordinates $x^\mu$. We assume the brane to be created at a
certain point in $B$, i.e. $Y^m(x)=Y^m_0$ which corresponds to its
ground state. The induced metric on the brane in such state is
given by the four-dimensional components of the bulk space metric,
i.e. $g_{\mu\nu}=\tilde g_{\mu\nu}=G_{\mu\nu}$. However, when
brane excitations (branons) are present, the induced metric is
given by
\begin{eqnarray}
g_{\mu\nu}=\partial_\mu Y^M\partial_\nu Y^N G_{MN} =\tilde
g_{\mu\nu}-\partial_{\mu}Y^m\partial_{\nu}Y^n\tilde g'_{mn}.
\end{eqnarray}
For illustrative purposes we show a toy model in Fig. 1 where we
have a 1-brane (string) in a $M_3=M_2\times S^1$ bulk-space
(one of the $M_2$ coordinates being the time coordinate), both
in its ground state (flat brane) and in an excited state (wavy
brane).

Since the mechanism responsible for the creation of the brane is
in principle unknown, we will assume that the brane dynamics can
be described by  an effective action. Thus, we will consider the
most general expression that is invariant under
reparametrizations of the brane coordinates. Following the
philosophy of the effective lagrangian technique, we will also
organize the action as a series in the number of the derivatives
of the induced metric over a dimensional constant, which can be
identified with the brane tension scale $f$. Therefore, up to
second order in derivatives we find:
\begin{equation}
S_B=\int_{M_4}d^4x\sqrt{g}\left( -f^4+\lambda f^2 R + \dots\right),
\label{Nambu4}
\end{equation}
where $d^4x\sqrt{g}$ is the volume element of the brane, $R$ the
induced curvature and $\lambda$ an unknown dimensionless
parameter. Notice that the lowest order term is the usual
Dirac-Nambu-Goto action that was the only one considered in
\cite{Dobado}. The brane-induced
curvature terms were first discussed in this context in \cite{Dvali}.
\begin{figure}
\vspace*{0cm}
\centerline{\mbox{\epsfysize=5.5 cm\epsfxsize=5.5
cm\epsfbox{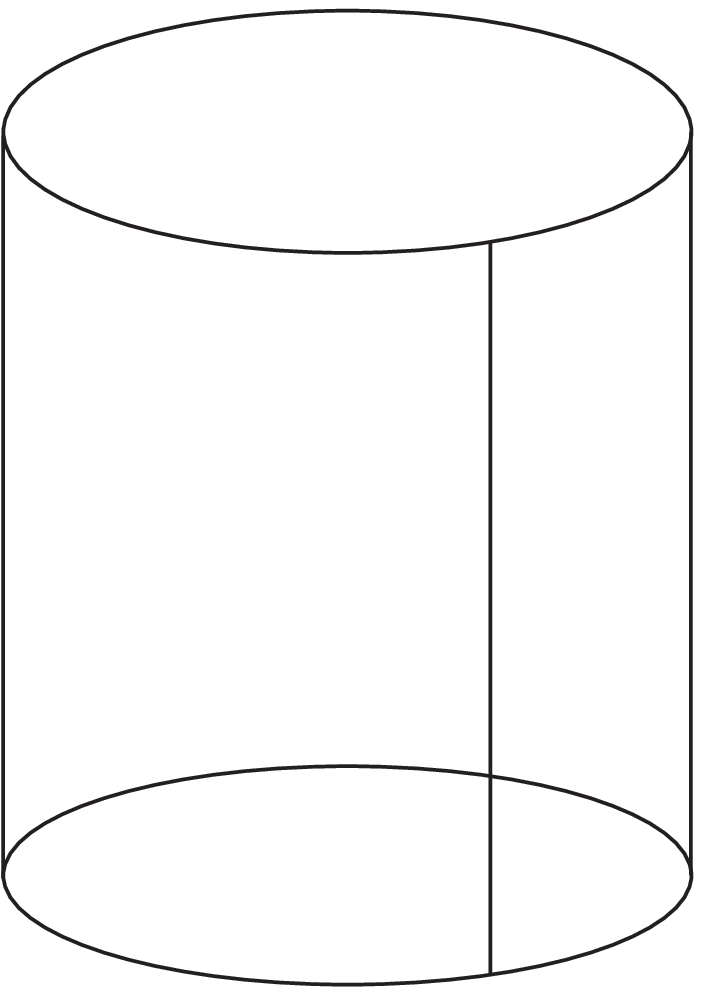}}\hspace*{2 cm}\mbox{\epsfysize=5.5
cm\epsfxsize=5.5 cm\epsfbox{cilmonel1.eps}}}
%
 \caption{\footnotesize{Brane with trivial topology in $M_3=M_2\times S^1$.
The ground
state of the brane is represented on the left. On the right we plot
an excited stated.}}
\end{figure}

As shown in \cite{Dobado}, if the brane ground state is 
$Y^m(x)=Y^m_0$, the presence of the
brane will break spontaneously all the $B$ isometries, except
those that leave the point $Y_0$ unchanged. In other words the
group $G(B)$ is spontaneously broken down to the isotropy group $H(Y_0)$
 of the point
$Y_0$. We will denote $\xi_\alpha$ the Killing fields  associated with the
broken generators of the group $G(B)$. The excitations of the brane 
along $\xi_\alpha$  correspond to the zero modes and
they are parametrized by the branon fields $\pi^\alpha(x)$ which 
can be understood as coordinates on the coset manifold $K=G(B)/H$.
Thus, for a position-independent ground state $Y^m_0$, the action
of an element of $G(B)$ on $B$ will map $Y_0$ into some other
point with coordinates:
\begin{equation}
Y^m(x)=Y^m(Y_0,\pi^\alpha(x))=Y^m_0+\frac{1}{k
f^2}\xi^m_\alpha(Y_0)\pi^\alpha(x)+\Od(\pi^2), \label{ypi1}
\end{equation}
where we have set the appropriate normalization of the branon
fields and Killing fields   with $k^2=16\pi /M_P^2$.  When
the $B$ space is homogeneous, the isotropy group does not depend
on the particular point we choose, i.e, $H(Y_0)=H$ and it is
possible to prove that $B$ is diffeomorphic to $K=G(B)/H$, i.e.
the number of GB equals the dimension of $B$. In that case we can
choose the coordinates on $B$ and $K$ so that
\begin{equation}
\pi^\alpha=\frac{v}{R_B}\delta_m^\alpha y^m=f^2\delta_m^\alpha y^m,
\end{equation}
or
\begin{equation}
Y^m=\frac{1}{f^2}\delta_\alpha ^m \pi^\alpha ,
\label{ypi2}
\end{equation}
where
\begin{equation}
v=f^2 R_B , \label{rel}
\end{equation}
is the typical size of $K$  in energy units (note that the
coordinates on $K$ must be scalar fields) and $R_B$ is the $B$
typical scale in length units. However, in the general
case, $B$ is not homogeneous and the number of GB will be smaller
than the dimension of $B$.
Notice that since $\pi^\alpha$ are
properly normalized scalar fields, the $Y^m$ coordinates in
(\ref{ypi2}) must be normal and geodesic in a neighborhood of $Y^m_0$
and, in particular, they cannot be angular coordinates.

According to the previous discussion, we can write the induced metric 
in terms of branon fields, thus using (\ref{ypi1}) we get:
\begin{equation}
g_{\mu\nu}=\tilde
g_{\mu\nu}-\frac{1}{f^4}h_{\alpha\beta}(\pi)\partial_{\mu}\pi^\alpha
\partial_{\nu}\pi^\beta
\end{equation}
where $h_{\alpha\beta}(\pi)$ is defined as
\begin{equation}
h_{\alpha\beta}(\pi)=f^4 \tilde g'_{mn}(Y(\pi))\frac{\partial
Y^m}{\partial\pi^\alpha}\frac{\partial Y^n}{\partial\pi^\beta}.
\end{equation}
Using these results, it is also  possible to obtain the 
expansion of the effective action in (\ref{Nambu4}) in terms
of branons (for a detailed discussion see the Appendix).

\section{Ground state of the brane and branon masses}

In the previous section we assumed that the $G(M_D)$ symmetry
is exact, which implies that the branon fields are massless.
However, in a real situation, such symmetry will be only
approximately realized. In this case, we will expect the branons
to acquire a mass that will measure the breaking of the $G(M_D)$
symmetry. In order to study these effects and how the symmetry
breaking affects the brane ground state,
 we will
relax the conditions imposed in the previous section
and let $\tilde g_{\mu\nu}$ depend, not only on the $x$ coordinates, but
also on the $y$ coordinates. We can consider
for simplicity the lowest-order action, given by:
\begin{eqnarray}
S_{eff}^{(0)}[\pi]&=&\int_{M_4}d^4x {\mathcal L}^{(0)}= -f^4
\int_{M_4}d^4x\sqrt{\tilde g(x,Y(x))}.
\end{eqnarray}
This action has an extremum if
\begin{eqnarray}
\delta S_{eff}^{(0)}[\pi]=0\Rightarrow \delta\sqrt{\tilde
g}=\frac{1}{2}\sqrt{\tilde g}\tilde g^{\mu\nu}\delta \tilde
g_{\mu\nu}=0\Rightarrow \tilde g^{\mu\nu}\partial_{m} \tilde
g_{\mu\nu}=0, \forall   y^m.
\end{eqnarray}
This is a set of equations whose solution $Y^m_0(x)$ determines
the shape of the brane in its ground state for a given background
metric $\tilde g_{\mu\nu}$.
In addition, the condition for the energy to be a minimum requires:
\begin{eqnarray}
\left.\frac{\delta^2 {\mathcal L}^{(0)}}{\delta Y^m\delta
Y^n}\right|_{Y=Y_0}&=&\frac{-f^4}{4}\sqrt{\tilde g}\tilde
g^{\mu\nu}(\partial_{n}\partial_{m} \tilde g_{\mu\nu}-2\tilde
g^{\rho\sigma}\partial_{n} \tilde g_{\nu\sigma}\partial_{m} \tilde
g_{\mu\rho})<0. \label{condsy}
\end{eqnarray}
i.e. the eigenvalues of the above matrix should be negative. This
implies that the static action should have a maximum (Dashen
condition). If we only focus on the degrees of freedom associated
to the branons, the previous conditions take the form:
\begin{eqnarray}
\frac{\delta {\mathcal L}^{(0)}}{\delta
\pi^\alpha}&=&\frac{-f^2}{2k}\sqrt{\tilde g}\tilde
g^{\mu\nu}\partial_{m} \tilde g_{\mu\nu}\xi^m_\alpha=0, \label{condspi}\\
\frac{\delta^2 {\mathcal L}^{(0)}}{\delta \pi^\alpha\delta
\pi^\beta}&=&-\frac{1}{4k^2}\sqrt{\tilde g}\tilde
g^{\mu\nu}(\partial_{n}\partial_{m} \tilde g_{\mu\nu}-2\tilde
g^{\rho\sigma}\partial_{n} \tilde g_{\nu\sigma}\partial_{m} \tilde
g_{\mu\rho})\xi^m_\alpha\xi^n_\beta < 0. \nonumber
\end{eqnarray}

In order to obtain the explicit expression for the branon mass
matrix, let us consider the following simple case:
\begin{eqnarray}
 G_{MN}&=&
\left(
\begin{array}{cccc}
\tilde g_{\mu\nu}(x,y)&0\\ 0&-\tilde g'_{mn}(y)
\end{array}\right),
\end{eqnarray}
where again $\tilde g_{\mu\nu}(x,Y_0)$ corresponds to the ground
state metric in the symmetric case. Expanding this metric around
$y^m=Y^m_0$ in terms of the $\pi^\alpha$ fields and taking into account 
the conditions for the ground state of the
brane (\ref{condspi}), the effective action is then given by
\begin{eqnarray}
S_B &=&-f^4 \int_{M_4}d^4x\sqrt{\tilde g}  \nonumber \\
&+&\frac{1}{2}\int_{M_4}d^4x\sqrt{\tilde g} (\tilde
g^{\mu\nu}h_{\alpha\beta}(\pi)\partial_{\mu}\pi^\alpha\partial_{\nu}\pi^\beta
-M^2_{\alpha\beta}
\pi^\alpha\pi^\beta)+...
\end{eqnarray}
where $\tilde g^{\mu\nu}$ in the previous equation
denotes $\tilde g^{\mu\nu}(x,Y_0)$ and
the mass matrix can be written as
\begin{equation}
M^2_{\alpha\beta}=\frac{1}{2}\tilde
g^{\mu\nu}(\partial_{n}\partial_{m} \tilde g_{\mu\nu}-2\tilde
g^{\rho\sigma}\partial_{n} \tilde g_{\nu\sigma}\partial_{m} \tilde
g_{\mu\rho})\frac{\xi^m_\alpha\xi^n_\beta}{k^2} > 0,
\end{equation}
which must be positive definite, otherwise the brane ground state
would not be stable. As an example, and for further reference let
us consider a background metric given by:
\begin{equation}
\tilde g_{\mu\nu}=(1+\sigma(y^2))\eta_{\mu\nu},
\label{massej}
\end{equation}
where $y_m$ are normal geodesic coordinates on $B=S^3$,
$y^2=\sum_m y_m^2$ and the function $\sigma(y^2)$ reaches its
minimum value at $y=0$ with $\sigma(0)=0$. The ground state
conditions imply:
\begin{eqnarray}
\partial_{m} \sigma\frac{\xi^m_\alpha}{k}=0
\end{eqnarray}
and
\begin{equation}
M^2_{\alpha\beta}=2\partial_{n}\partial_{m}
\sigma\frac{\xi^m_\alpha\xi^n_\beta}{k^2} > 0.
\end{equation}
Since  in this case we have $\xi^m_\alpha=k\delta^m_\alpha$ we
obtain a diagonal mass matrix with all the branons having the same
mass $m^2=4\sigma'(0)$ and then:
\begin{equation}
M^2_{\alpha\beta}=m^2\delta_{m
n}\frac{\xi^m_\alpha\xi^n_\beta}{k^2}=m^2 \delta_{\alpha\beta}.
\end{equation}

\section{Brane-skyrmions}

The branon fields introduced in the previous sections describe
small oscillations of the brane around its ground state. Thus
there is some similarity with the well-known chiral lagrangian
approach in which a non-linear sigma model (NLSM) is used to
describe the low-energy pion dynamics. Apart from pions, the NLSM
can also be used to describe other non-trivial states in the
hadron spectrum such as baryons. For that purpose, the non-trivial
topological structure of the coset space $K$ plays a fundamental
role.  In fact, baryons can be identified with certain
topologically non-trivial maps between the (compactified) space
$S^3$ and the coset manifold $K$ known as Skyrmions.

Let us then consider static branon field configurations with
finite energy, which accordingly vanish at the spatial infinity.
Thus, we can compactify the spatial dimensions to $S^3$ and the
static configurations will be mappings: $\pi^\alpha:
S^3\rightarrow K$. Therefore, these mappings can be classified
according to the third homotopy group of $K$, i.e. $\pi_3(K)$. As
a consequence, mappings belonging to different non-trivial
homotopy classes cannot be deformed in a continuous fashion from
one to the other. This implies that such configurations cannot
evolve in time classically into the trivial vacuum $\pi=0$ and
therefore they are stable states. For the sake of simplicity we
will first consider the case in which we have $N=3$ extra dimensions
with $B=S^3$. In this case, since $B$ is an homogeneous space, we
have $K\sim B=S^3 \sim SU(2)$, i.e the coset manifold is also a
3-sphere. Thus, we will have $\pi_3(S^3)=\bf{Z}$ and the mappings
can be classified by an integer number, usually referred to as the
winding number $n_W$. We will also assume in the following, that
the background metric is flat, i.e, $\tilde g_{\mu\nu}=
\eta_{\mu\nu}$.

For static configurations the  brane-skyrmion mass can be
obtained directly from the effective lagrangian as:
\begin{eqnarray}
M[\pi]=-\int d^3 x {\cal L}_{eff}.
\end{eqnarray}
In general this expression will be divergent because of the volume
contribution coming from the lowest order term
$S^{(0)}_{eff}[\pi]$ which reflects the fact that the brane has
infinite extension with  finite tension. Therefore, in order to
obtain a finite skyrmion mass, we will subtract the vacuum energy
$M[0]$, i.e.:
\begin{equation}
M_S[\pi]=M[\pi]-M[0]=f^4 \int_{M_3}d^3x\sqrt{g}-\lambda f^2
\int_{M_3}d^3x\sqrt{g}R-M[0]. \label{mass}
\end{equation}
In other words we are defining the mass of the brane-skyrmion as
the mass of the brane with the topological defect minus the mass
of the brane in its ground state without topological defect.

In order to simplify the calculations we will introduce spherical
coordinates on both spaces, $M_4$ and $K$. In $M_4$ we denote the
coordinates $\{t,r,\theta,\varphi\}$ with $\phi \in [0,2\pi)$,
$\theta \in[0,\pi]$ and $r \in [0,\infty)$. In these coordinates
the background metric is written as:
\begin{eqnarray}
\tilde g_{\mu\nu}=\left(
\begin{array}{cccc}
1&\\ &-1&\\ & &-r^2&\\ & & & -r^2 \sin^2(\theta)
\end{array}\right).
\end{eqnarray}

On  the coset  manifold $K$, the spherical coordinates are denoted
$\{\chi_K,\theta_K,\phi_K\}$ with $\phi_K \in [0,2\pi)$, $\theta_K
\in [0,\pi]$ and $\chi_K \in [0,\pi]$. These coordinates are
related to the physical branon fields (local normal geodesic
coordinates on $K$) by:
\begin{eqnarray}
\pi_1&=&v \sin\chi_K\sin\theta_K\cos\phi_K,\nonumber\\
\pi_2&=&v \sin\chi_K\sin\theta_K\sin\phi_K,\label{esfk}\\ \pi_3&=&v
\sin\chi_K\cos\theta_K.\nonumber
\end{eqnarray}

The coset metric in spherical coordinates is written as
\begin{eqnarray}
h_{\alpha\beta}=\left(
\begin{array}{ccc}
v^2&\\ &v^2\sin^2(\chi_K) &\\ & & v^2
\sin^2(\chi_K)\sin^2(\theta_K)
\end{array}\right).
\end{eqnarray}

In spherical coordinates, the brane-skyrmion with winding number
$n_W$ is given by the non-trivial mapping
$\pi^\alpha:S^3\longrightarrow S^3$ defined from:
\begin{eqnarray}
\phi_K&=&\phi,\nonumber
\\ \theta_K&=&\theta,\nonumber\\
\chi_K&=&F(r),\label{skyan}
\end{eqnarray}
with boundary conditions $F(0)=n_W\pi$ and $F(\infty)=0$. This
map is usually referred to as the hedgehog ansatz. In Fig. 2 we
plot on the right a Skyrmion-like 1-brane configuration in
$M_3=M_2\times S^1$. In this particular dimension the
brane-skyrmion is not stable (as shown below), but still it is useful for
illustrative purposes.

\begin{figure}
\vspace*{0cm}
\centerline{\mbox{\epsfysize=5.5 cm\epsfxsize=5.5
cm\epsfbox{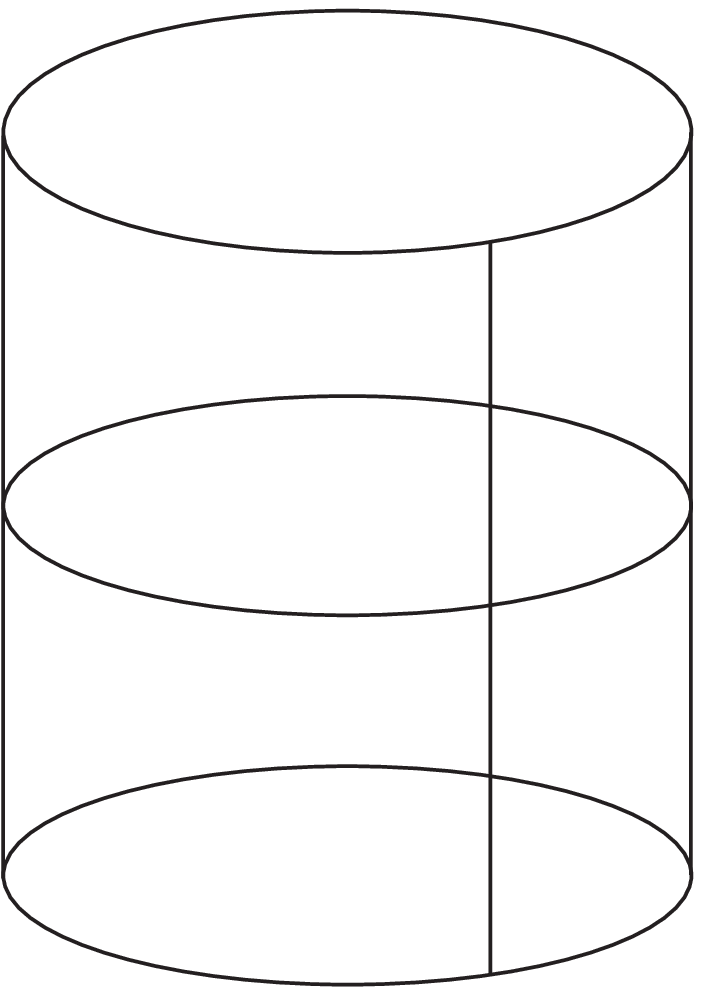}}\hspace*{2 cm}\mbox{\epsfysize=5.5
cm\epsfxsize=5.5 cm\epsfbox{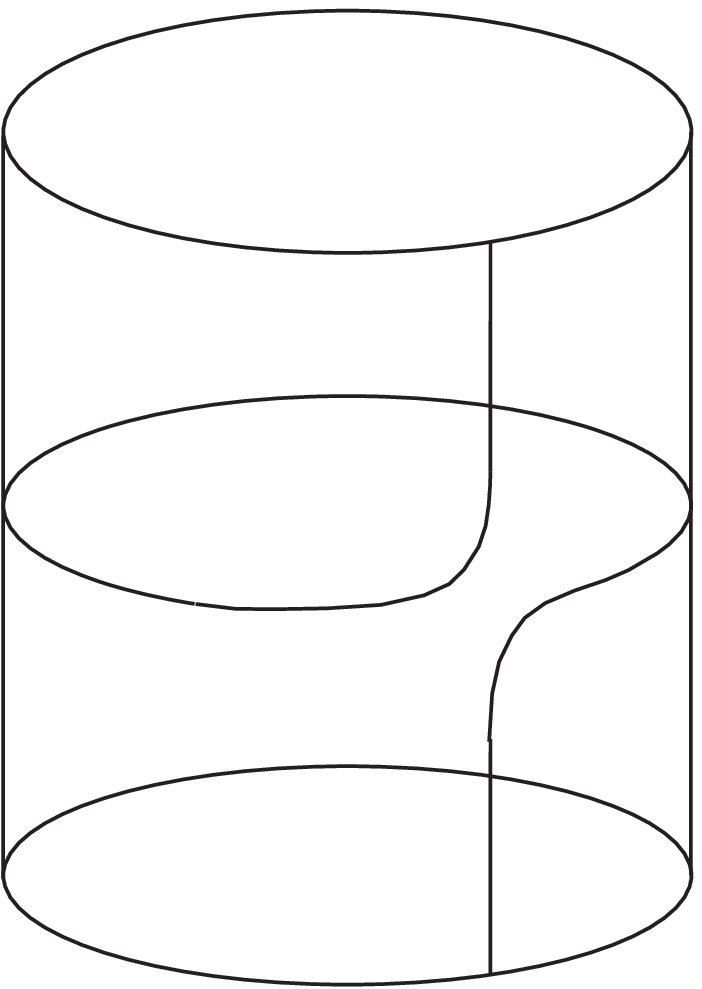}}}
%
 \caption{\footnotesize{Brane configurations with $n_W=1$
in $M_3=M_2\times S^1$. On the right we plot  a non-zero size
Skyrmion. On the left a  zero-size 1-brane-skyrmion  is shown. It
has the same mass and  shape as the state built out of a wrapped soliton 
and a topologically trivial (world) brane. However the topology is not
the same (see section 9).}}
\end{figure}
Notice once again
 that the spherical coordinates
we have introduced on $K$ cannot be directly understood as branon
fields, although, thanks to the reparametrization invariance of
the action, they are perfectly valid for performing the calculations.

In order to calculate the brane-skyrmion mass from (\ref{mass}),
we need explicit expressions for the induced metric determinant
and the scalar curvature. In these coordinates the induced metric
on the brane is given by the diagonal matrix
\begin{eqnarray}
g_{\mu\nu}=\left(
\begin{array}{cccc}
1\vspace{.1cm}&\\
&\hspace{-.4cm}-\left(1+\frac{v^2}{f^4}(F'(r))^2\right)\vspace{.1cm}&\\
& &\hspace{-1.7cm}-\left(r^2+\frac{v^2}{f^4}
\sin^2(F(r))\right)\vspace{.1cm}&\\ & & &
\hspace{-2cm}-\left(r^2+\frac{v^2}{f^4}
\sin^2(F(r))\right)\sin^2(\theta)
\end{array}\right),
\end{eqnarray}
and the corresponding square root of the determinant 
can be easily found to be:
\begin{eqnarray}
\sqrt{g}=\sqrt{1+\frac{v^2}{f^4}(F'(r))^2}\left(r^2+\frac{v^2}{f^4}
\sin^2(F(r))\right) \sin(\theta).
\end{eqnarray}
The induced scalar curvature can be obtained in a simple way by
introducing the isotropic coordinates
$\{t,\rho,\theta,\varphi\}$ where $\rho$ is defined as
\begin{equation}
\rho^2\equiv r^2+\frac{v^2}{f^4} \sin^2(F(r)).\label{rho}
\end{equation}
The metric in these new coordinates is written in the isotropic
canonical form:
\begin{eqnarray}
g_{\mu\nu}=\left(
\begin{array}{cccc}
1&\\ &-A(r)&\\ & &-\rho^2&\\ & & & -\rho^2 \sin^2(\theta)
\end{array}\right),
\end{eqnarray}
where
\begin{eqnarray}
A(r)=\frac{1+\frac{v^2}{f^4}(F'(r))^2}{\rho'^2},\label{A}
\end{eqnarray}
and  $'$ denotes $r$ derivative. The scalar curvature for this
metric turns out to be:
\begin{eqnarray}
R=-\frac{2}{\rho^2}\left(1-\frac{1}{A(r)}\right)-\frac{2A'(r)}{A^2(r)\rho\rho'}.
\end{eqnarray}

Thus we can write $M_S$ as a functional of $F(r)$. The actual mass
of the brane-skyrmion with winding number $n_W$ will be obtained
by minimizing the functional $M_S[F]$ (\ref{mass}) in the space
of functions $F(r)$ with the appropriate boundary conditions. This problem is
in general rather complicated, but we can obtain an upper bound to
the mass by using a family of functions parametrized by a single
parameter, and minimizing with respect to that parameter. In
particular, it is very useful to work with the Atiyah-Manton
ansatz \cite{Atiyah}:
\begin{eqnarray}
 F(r)=n_W \pi\left(1-\frac{1}{\sqrt{1+L^2/r^2}}\right).
\label{AM}
\end{eqnarray}
By minimizing the brane-skyrmion mass with
respect to $L$ we will get: $M_S\equiv min_L\; M_S(L)\equiv M_S(L_m)$
for different values of the parameter $\lambda$. Thus:
\begin{itemize}
\item For
$\lambda=0$ and using the ansatz above for $n_W=1$ we find that
$M_S(L)$ is minimized for $L_m=0$. The corresponding mass is given
by
\begin{equation}
M_S=2\pi^2f^4R_B^3,
\end{equation}
i.e. the brane-skyrmion describes a
pointlike particle with a finite mass given by the volume of the
extra dimensions times the brane tension $f^4$. It can be shown
(see next section) that this result is general, i.e. it is valid
for any parametrization of the $F(r)$ function and not only for
the Atiyah-Manton one.

\item For $\lambda>0$, we find that $M_S(L)$ is minimized for some $L_m>0$
(non-zero size brane-skyrmion). In this case, assuming that
 the contribution
from the curvature term never becomes negative, it is
possible to obtain the following lower bound on the mass
$M_S>M_S(\lambda=0)=2\pi^2f^4R_B^3$. An upper bound can be
obtained by  evaluating $M_S$ in the limit $L=0$,
simply assuming that this limit is well defined so that
we can commute the limit with the integration. Thus
\begin{equation}
M_S<M_S(L=0)=2\pi^2f^4R_B^3\left(1+6\frac{\lambda}{R_B^2f^2}\right).
\end{equation}
These results are general for any monotonic parametrization and, in particular,
we have checked numerically that they hold for the Atiyah-Manton case.

\item Finally, for $\lambda<0$, we find again that the minimum $M_S$
corresponds to zero size brane-Skyrmion and the corresponding mass
is
\begin{equation}
M_S=2\pi^2f^4R_B^3\left(1 +6\frac{\lambda}{R_B^2f^2}\right).
\end{equation}
When $\lambda<-R_B^2f^2/6$ the brane-skyrmion mass becomes
negative. In this case, using non-monotonic parametrizations, we
have obtained that the mass  is actually not bounded from below,
since the curvature term can be made arbitrarily large. As a
consequence only in this last case, the brane-skyrmion becomes
unstable. These results are summarized in Table.1
\end{itemize}
\begin{table}[h]
\begin{center}
\begin{tabular}{|c|c|c|}
\hline
 & & \\
 $\lambda$& Size &   Mass \\
 & & \\
\hline $\lambda=0$& $L_m=0$ & $M_S=2\pi^2f^4R_B^3$\\ \hline
$\lambda>0$ &$L_m>0$ & $2\pi^2f^4R_B^3<M_S<2\pi^2f^4R_B^3(1
+6\frac{\lambda}{R_B^2f^2})$ \\ \hline $-R_B^2f^2/6<\lambda<0$ &
$L_m=0$ & $M_S=2\pi^2f^4R_B^3(1 +6\frac{\lambda}{R_B^2f^2})$ \\
\hline $\lambda<-R_B^2f^2/6$ & $L_m=0$ & $M_S\rightarrow
-\infty$\\ \hline
\end{tabular}
\caption{\footnotesize{Values of the size $L_m$ and mass $M_S$ for
the brane-skyrmion with $n_W=1$ for different values of the
$\lambda$ parameter}}
\end{center}\label{masas}
\end{table}
In Fig.3 we show the brane-skyrmion mass as a function of
$\lambda$ and $L$. We see that the minimum is displaced from $L=0$
when $\lambda>0$.
\begin{figure}[h]
\vspace*{0cm}
\centerline{\mbox{\epsfysize=10 cm\epsfxsize=10
cm\epsfbox{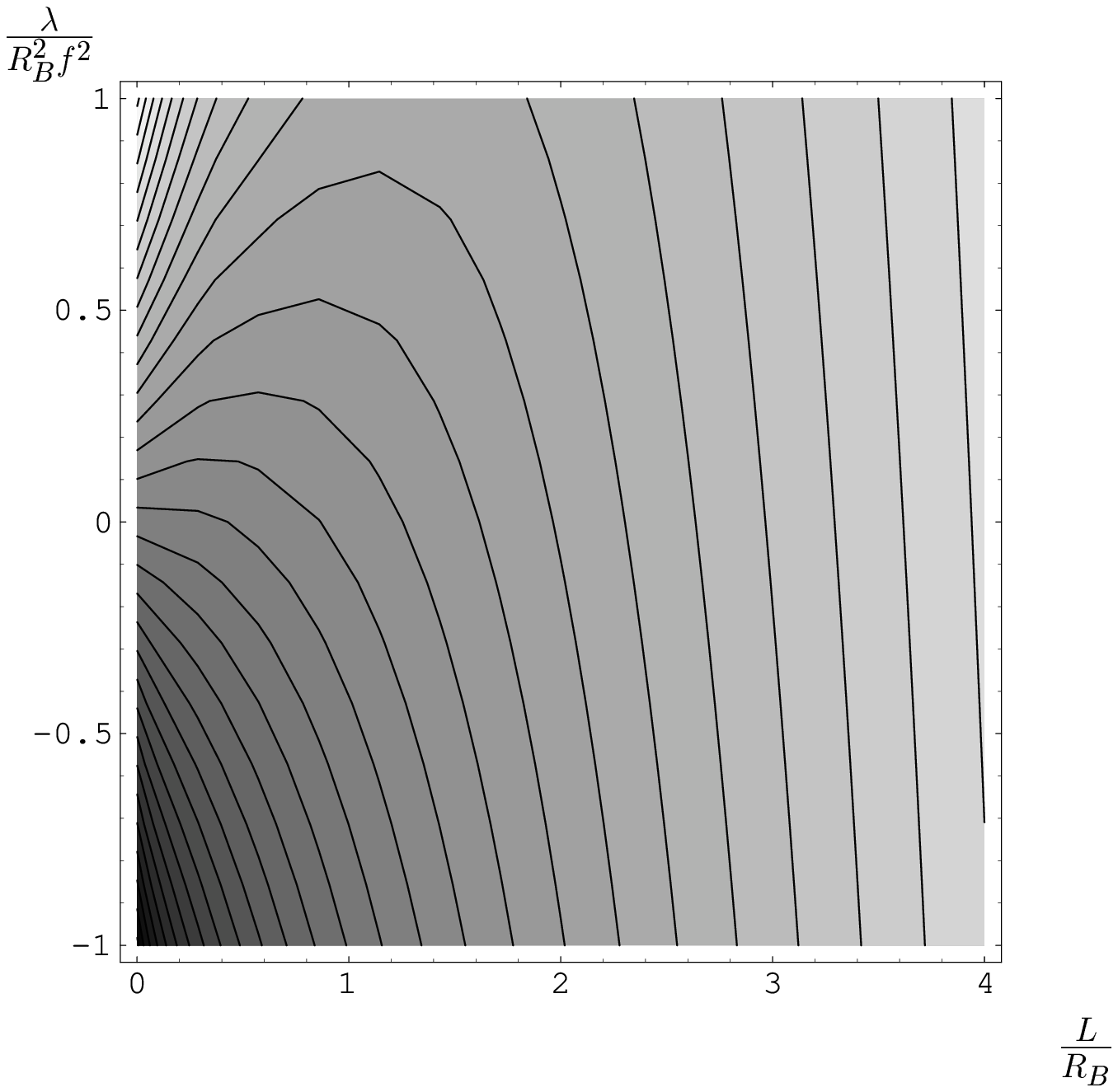}}}
%
 \caption{\footnotesize{Contour plot of the brane-skyrmion mass as a
function of
 $\lambda/(R_B^2f^2)$ and $L/R_B$.}}
\end{figure}

Notice that these results differ a lot from those obtained in the
usual $SU(2)$ Skyrme model in QCD chiral dynamics \cite{Skyrme,Witten1}.
That model is
based on a fourth-order lagrangian and only within a certain range
of the parameters $M$ and $N$ of the two independent
four-derivative terms, the Skyrmion can be made stable. In our
case, the lagrangian contains an infinite number of terms, which
are responsible for the brane-skyrmion stability. In fact,
truncating the series and keeping only up to fourth order terms,
one obtains in our case:
\begin{eqnarray}
 S_{eff}^{(2,4)}[\pi]&=&\frac{1}{2}\int_{M_4}d^4x\sqrt{\tilde g}
h_{\alpha\beta}\partial_{\mu}\pi^\alpha\partial^{\mu}\pi^\beta
\nonumber\\ &+&\frac{1}{4v^4}\int_{M_4}d^4x\sqrt{\tilde
g}h_{\alpha\beta}
h_{\gamma\delta}(M\partial_{\mu}\pi^\alpha\partial^{\mu}
\pi^\beta\partial_{\nu}\pi^\gamma\partial^{\nu}\pi^\delta\nonumber\\
&+&N\partial_{\mu}\pi^\alpha\partial^{\nu}\pi^\beta\partial_{\nu}
\pi^\gamma\partial^{\mu}\pi^\delta),
\end{eqnarray}
where we have assumed $\tilde g_{\mu\nu}=\eta_{\mu\nu}$. The
corresponding  $M$ and $N$ parameters in the chiral lagrangian are
then given by:
\begin{equation}
M=\frac{-f^4R_B^4}{32}, \;\; N=\frac{f^4R_B^4}{16}. \label{MN}
\end{equation}
This implies that the standard $e^2$ and $\gamma$ Skyrme-model
parameters are
\begin{equation}
e^2=\frac{1}{16N}=\frac{1}{f^4R_B^4}>0, \;\; \gamma=\frac{1}{2}
\left(1+\frac{M}{N}\right)=\frac{1}{4}>0,
\end{equation}
which means that the Skyrmion would not be stable if one had
considered the truncated fourth-order lagrangian only. However, for
the full lagrangian (with $\lambda>-R_B^2f^2/6$) we have seen
before that the brane-skyrmion is stable. This shows that the infinite number
of higher-order terms are responsible for its stability. This of course
could also be the case in chiral dynamics. In this case, it is also 
possible to
compute the $M$ and $N$ parameters in the framework of large $N_c$
Quantum Cromodynamics (QCD) (for a review see \cite{Book})
\begin{equation}
M=\frac{N_c}{192\pi^2}, \;\; N=\frac{-N_c}{384\pi^2},
\end{equation}
which implies $M/N=-2$. It is quite amazing to realize that this is exactly
the same relation found in the branon dynamics. In particular for:
\begin{equation}
N_c=12\pi^2f^4R_B^4
\end{equation}
we have the same answer in the two models, provided we identify
$v=F_{\pi}$, $F_{\pi}$ being the pion decay constant. Thus for $N_c=3$ one
can set $R_B^2=1/(2\pi f^2)$ in such a way that the branon
dynamics could reproduce  the correct low-energy  pion scattering
and give rise to a stable nucleon. We really do not have any
particular reason to understand why this simple brane model works 
so well apart from having the same symmetry breaking pattern as
the low-energy chiral dynamics.

\section{Numerical results}

As we saw in the previous section, if $\lambda>0$, the
brane-skyrmion size is different from zero and its mass cannot be
evaluated analytically although some bounds can be obtained. For
this reason, it is very interesting to obtain the behaviour of the
brane-skyrmion size and  mass as a function of $\lambda$, at least
numerically. As shown in the previous section, the contribution of
the curvature term to the brane-skyrmion mass depends on
$\lambda/(R_B^2f^2)$. Assuming that such contribution is small we
have performed a numerical analysis in the case $B=S^3$ using the
Atiyah-Manton parametrization. For the brane-skyrmion size in the
mentioned limit we find the following behaviour:
\begin{eqnarray}
\frac{L_m}{R_B}=\left\{\begin{array}{cc}0;&\lambda\leq 0\\ & \\
a\left(\frac{\lambda}{R_B^2f^2}\right)^\alpha;&\lambda>0\end{array}\right. ,
\end{eqnarray}
where the value of the  $a$ and $\alpha$  parameters are given in
Table.2. In Figure 4, we show this non-analytic behavior for the
brane-skyrmion size with $n_W=1$.
\begin{figure}[h]
\vspace*{-3.5cm}
\centerline{\mbox{\epsfysize=10 cm\epsfxsize=10
cm\epsfbox{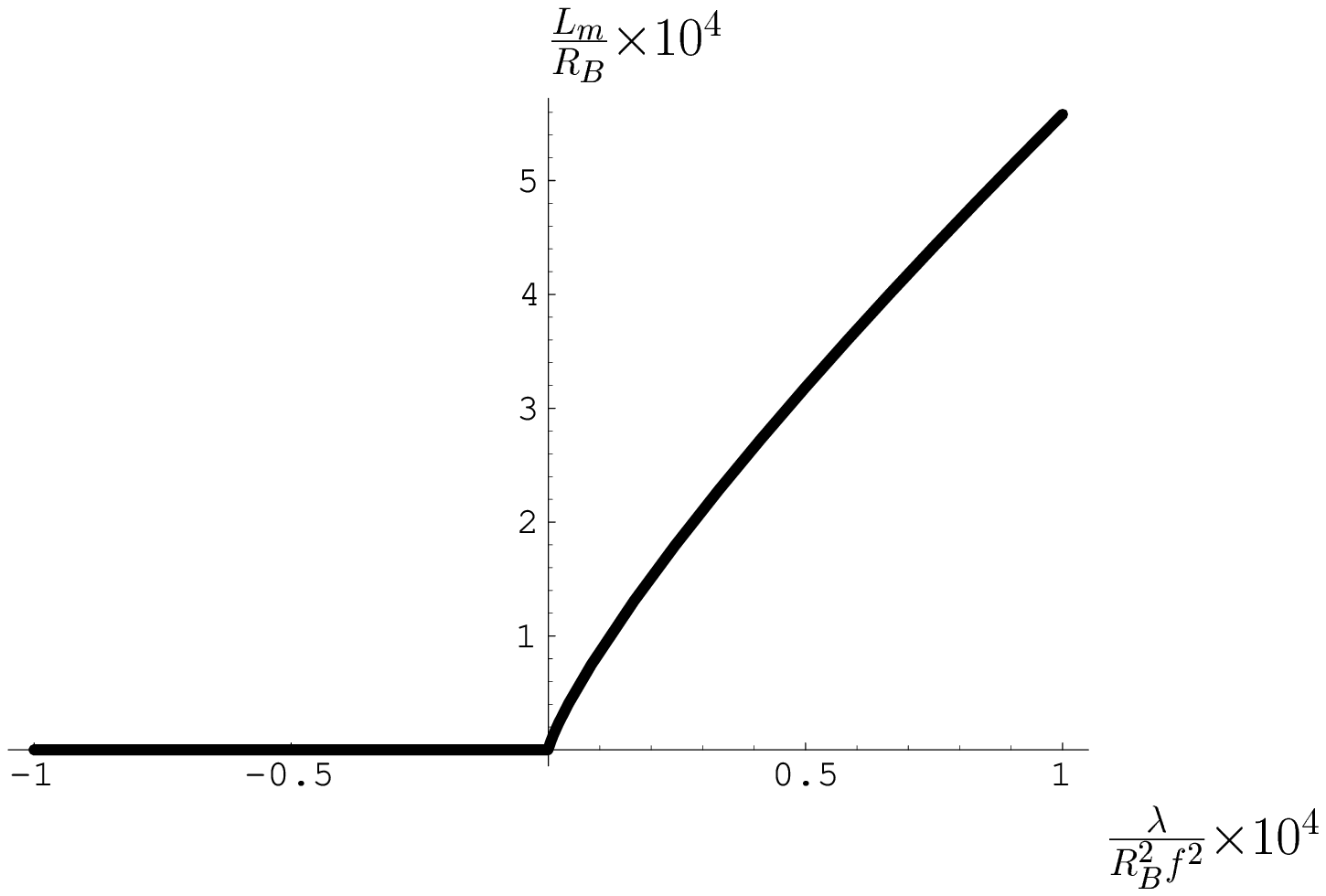}}}
%
 \caption{\footnotesize{Behaviour of the brane-skyrmion size with $n_W=1$ when
$\lambda/R_B^2f^2$ is small.}}
\end{figure}
The fact that the brane-skyrmion acquires a non-zero size implies
that its mass depends on $\lambda$ in a non-trivial way. The
numerical calculation shows that the behaviour of the
brane-skyrmion mass in the mentioned limit is of the form:
\begin{eqnarray}
\frac{M_S}{4\pi f^4R_B^3}=\left\{\begin{array}{cc}\frac{\vert
n_W\vert \pi}{2} +3\vert n_W\vert
\left(\frac{\lambda}{R_B^2f^2}\right);&\lambda\leq 0\\ &
\\ \frac{\vert n_W\vert \pi}{2}+3\vert n_W\vert
\left(\frac{\lambda}{R_B^2f^2}\right)
-b\left(\frac{\lambda}{R_B^2f^2}\right)^\beta;
&\lambda>0\end{array}\right. ,
\end{eqnarray}
where the value of the parameters $b$ and $\beta$ is shown in
Table.2 for different winding numbers $n_W$. The non-analytic
behaviour of the derivative with respect to $\lambda$ is plotted
in Fig.5 for $n_W=1$. The numerical calculation for the
Atiyah-Manton ansatz shows that the constants $a$ and $b$ depend on
$n_W$ in a non-trivial way, whereas the exponents $\alpha$ and
$\beta$ do not.
\begin{figure}[h]
\vspace*{-3.5cm}
\centerline{\mbox{\epsfysize=10 cm\epsfxsize=10
cm\epsfbox{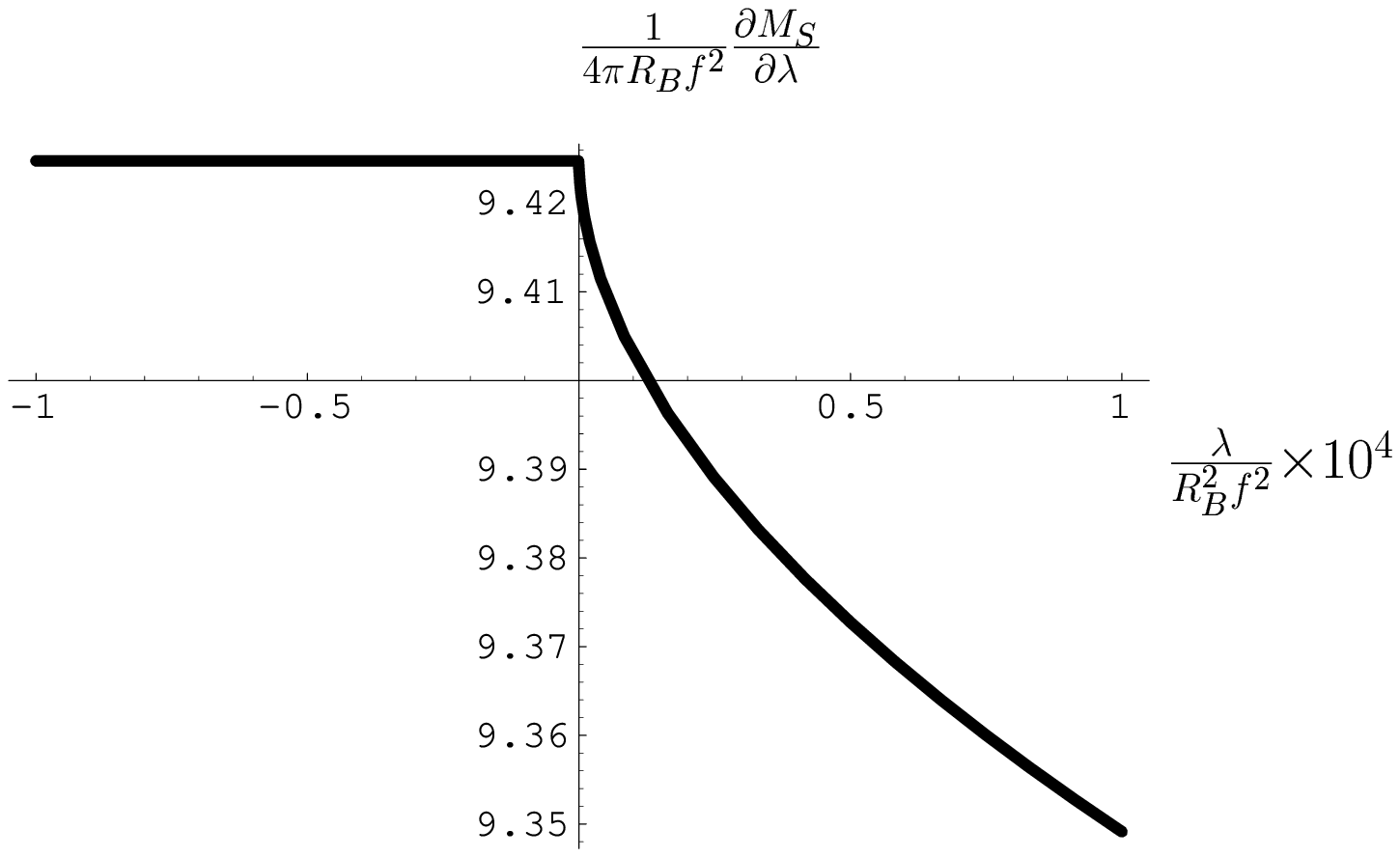}}}
%
 \caption{\footnotesize{Behaviour of the derivative with respect to
$\lambda$ of
 the brane-skyrmion mass with $n_W=1$ for small $\lambda/(R_B^2f^2)$.}}
\end{figure}
An interesting conclusion that can be obtained from these results
is that the interaction between two classical brane-skyrmions with
topological number $n_W=1$ is repulsive when their sizes are non
zero, whereas they do not interact if their sizes are exactly
zero. The reason is that one could understand a $n_W=2$
brane-skyrmion as two $n_W=1$ brane-skyrmions located at the same
point. The energy of the $n_W=2$ brane-skyrmion  is bigger than
the energy of two brane-skyrmions with $n_W=1$ if their sizes are
non zero, whereas it is the same if their sizes are exactly zero.
%
\begin{table}[h]
\begin{center}
\begin{tabular}{|c|cccc|}
\hline
Topological &$\alpha$&  $\beta$&$a$& $b$ \\ charge $n_W$& & & &\\
\hline
1&0.81&1.54&0.97&7.1\\
2&0.81&1.54&0.59&6.5\\
3&0.81&1.54&0.44&6.2\\
\hline
\end{tabular}
\caption{\footnotesize{Values of the constants $a$, $b$, $\alpha$
and $\beta$ for different topological numbers $n_W$ in the
Atiyah-Manton ansatz.}}
\end{center}
\end{table}

It is important to notice that these numerical results have been
obtained with the hedgehog ansatz and the Atiyah-Manton
parametrization and therefore they are only estimations of the
true mass and size. In fact, it is known from the standard Skyrme
model for baryons that the hedgehog ansatz is not appropriate for
the description of the deuteron, corresponding to a $n_W=2$
Skyrmion since there is other toroidal-like configuration with
less energy in this topological sector.

\section{Brane-skyrmions in higher dimensions}

In this section we will extend our previous discussion to an
arbitrary number $N$ of space and compactified dimensions. We will
start from the simplest case with $\lambda=0$. Let us consider the
Dirac-Nambu-Goto action for a N-brane:

\begin{equation}
S_B=-f^{N+1} \int_{M_{N+1}}dt\;d^N x\sqrt{g}. \label{NambuN}
\end{equation}
For simplicity, we will take $\tilde g_{\mu\nu}=\eta_{\mu\nu}$ and
$K=S^N$, so that $M_D=M_{N+1}\times S^N$ with $D=2N+1$. The
generalization of the hedgehog ansatz to $N$ dimensions with
winding number $n_W$ can be written more easily in angular
coordinates  as follows:
\begin{eqnarray}
\begin{array}{lll}
\phi_K=\phi & \phi_K \in [0,2\pi), & \phi \in [0,2\pi)\nonumber \\
\theta^i_K=\theta^i& \theta^i_K\in[0,\pi],& \theta^i\in[0,\pi], \;
 i=1..\;N-2
\nonumber \\ \chi_K=F(r) & \chi_K\in[0,n_W\pi],&
r\in[0,\infty) ,
\end{array}
\end{eqnarray}
where the $K$ subindex refers to coordinates on the $K$ manifold,
the angular coordinates on $M_{N+1}$ do not carry any subindex and
the chiral angle function $F(r)$ satisfies the boundary conditions
$F(0)=n_W\pi$, $F(\infty)=0$. Introducing again  $\rho$ as in (\ref{rho}),
the induced metric takes the spherically symmetric form:
\begin{eqnarray}
g_{\mu\nu}=\left(
\begin{array}{ccccccccc}
1&\\ &\hspace{-.3cm}-A(r)&\\ & &\hspace{-.3cm}-\rho^2&\\ & & &
\hspace{-.6cm}-\rho^2 \sin^2 \theta_{N-2} &\\ & & & &
\hspace{-.6cm}-\rho^2 \sin^2 \theta_{N-2}\sin^2 \theta_{N-3}&\\ &
& & & & \hspace{-3cm}. &\\ & & & & & & \hspace{-3cm}. &\\ & & & &
& & & \hspace{-3cm}. &\\ & & & & & & & & \hspace{-2.6cm}-\rho^2
\sin^2 \theta_{N-2}...\sin^2 \theta_{1}
\end{array}\right)
\end{eqnarray}
where $A(r)$ is given in equation (\ref{A}). For static
configurations, the Dirac-Nambu-Goto action (\ref{NambuN})
provides the mass functional:
\begin{eqnarray}
M_S&=&f^{N+1}\int d^N x \left(\det\left(
g_{\mu\nu}\right)^{1/2}-\det(\eta_{\mu\nu})^{1/2}\right) \nonumber
\\ &=&f^{N+1}\Omega_N\int dr \left(\left[r^2+\frac{v^2}{f^4}
\sin^2 F(r)\right]^{\frac{N-1}{2}}\left[1+\frac{v^2}{f^4}
F'(r)^2\right]^{\frac{1}{2}}-r^{N-1}\right),\nonumber \\
\label{Nmass}
\end{eqnarray}
where the total solid angle is given by:
\begin{eqnarray}
\Omega_N=\frac{2\pi^{N/2}}{\Gamma(N/2)}.
\end{eqnarray}
For $N\geq 2$ it is possible to obtain a non-trivial bound for
$M_S$. $N=1$ is a special case and will be studied later on. Let
us start from the inequality:
\begin{eqnarray}
(a+b)^{\frac{N-1}{2}}\geq (a^{N-1}+b^{N-1})^{1/2},
\end{eqnarray}
which is valid for $N\geq 2$ and $a,b>0$. We will also use:
\begin{eqnarray}
(a+b)^{1/2}(c+d)^{1/2}\geq ac+bd,
\end{eqnarray}
again for $a,b,c,d>0$. From (\ref{Nmass}) we obtain:
\begin{eqnarray}
M_S\geq f^{N+1}\Omega_N \int dr \left(\frac{v}{f^2}\right)^N
\vert\sin^{N-1}(F(r))\vert \vert F'(r)\vert .
\end{eqnarray}
Since $v=R_Bf^2$ we have:
\begin{eqnarray}
M_S &\geq& f^{N+1}R_B^N\Omega_N \int dr \vert\sin^{N-1}(F(r))\vert
\vert F'(r)\vert \nonumber \\ &=&f^{N+1}R_B^N\Omega_N
\int_0^{n_W\pi} du \vert\sin^{N-1}(u)\vert ,
\end{eqnarray}
where in the last step, we have defined $u=F(r)$, assuming that
$F(r)$ is a monotonic function.

The total volume of the $S^N$ sphere with radius $R_B$ is given
by:
\begin{eqnarray}
V_N(R_B)=\frac{2\pi^{\frac{N+1}{2}}}{\Gamma\left(\frac{N+1}{2}\right)}
R_B^N=R_B^N\Omega_{N+1} .
\end{eqnarray}
Accordingly, we can write the bound on the mass as:
\begin{eqnarray}
M_S\geq M_{Sm}\equiv \vert n_W\vert f^{N+1}R_B^N\Omega_{N+1}
=\vert n_W\vert f^{N+1}V_N(R_B).
\end{eqnarray}
This is a lower bound for $N\geq 2$. On the other hand, we can
write the chiral angle as $F(r)=\bar F(r/L)=\bar F(\bar r)$ with
$\bar r=r/L$, $L$ being the typical brane-skyrmion size. Thus, the
mass functional can be written as:
\begin{eqnarray}
M_S &=&f^{N+1}\Omega_N L\int d\bar r \left(\left[L^2\bar r^2+R_B^2
\sin^2 \bar F(\bar
r)\right]^{\frac{N-1}{2}}\left[1+\frac{R_B^2}{L^2}
\left(\frac{d\bar F}{d\bar r}\right)^2\right]^{\frac{1}{2}}\right.\nonumber \\
&-&\left.(L\bar
r)^{N-1}\right) .\label{NMassL}
\end{eqnarray}
Taking the limit $L\rightarrow 0$ in the above integral, we find
that it is well defined and $M_S\rightarrow M_{Sm}$, i.e., zero
size brane-skyrmions saturate the bound. Therefore the Skyrmion
mass is $M_S=M_{Sm}$ and the Skyrmion is pointlike ($L_m=0$) for
$N\geq 2$.

In the case $N=1$, the inequalities above cannot be used  and the
only bound is the trivial one, i.e. $M_S\geq 0$. It is possible to
show from (\ref{NMassL}) that in this case, the bound is saturated
in the opposite limit $L\rightarrow \infty$. Accordingly,
1-brane-Skyrmions in one extra dimension would be massless and
non-localized.

Let us consider now the effect of the brane-induced curvature $R$
term, i.e. $\lambda\neq 0$. This curvature is given by
\begin{eqnarray}
R=-(N-1)\left[\frac{(N-2)}{\rho^2}\left(1-\frac{1}{A(r)}\right)
+\frac{A'(r)}{A^2(r)\rho\rho'}\right].
\end{eqnarray}
Following the same steps as in the $N=3$ case, we obtain
 the following results. For $\lambda\in(0,\infty)$ and $n_W=1$ we
have $L_m>0$ and the mass of the brane-skyrmion is bounded in the range 

\begin{equation}
\frac{2\pi^\frac{N+1}{2}}{\Gamma(\frac{N+1}{2})}f^{N+1}R_B^N
<M_S<2\frac{2\pi^\frac{N+1}{2}}{\Gamma(\frac{N+1}{2})}
f^{N+1}R_B^N\left(1+N(N-1)\frac{\lambda}
{R_B^2f^2}\right).
\end{equation}

For $\lambda\leq 0$ the brane-skyrmion colapses i.e. $L_m=0$ and
the mass of the brane-skyrmion can be computed as in the $N=3$
case to find:

\begin{equation}
M_S=2\frac{2\pi^\frac{N+1}{2}}{\Gamma(\frac{N+1}{2})}f^{N+1}R_B^N\left(1+N(N-1)
\frac{\lambda}{R_B^2f^2}\right).
\end{equation}

Finally, for $\lambda<-R_B^2f^2/N(N-1)$ the brane-skyrmion mass is
negative and not bounded from below.

The previous generalization is appropriate for $N \geq 3$.  For
the particular case $N=2$ it is possible to show that the
curvature integrates to zero. This is because in this case we have
\begin{eqnarray}
R=-\frac{A'(r)}{A^2(r)\rho\rho'}
\end{eqnarray}
and
\begin{eqnarray}
g=A(r)\rho^2(\rho')^2.
\end{eqnarray}
Then, taking into account that $1/\sqrt{A}$ vanishes  whenever
$\rho'$ is zero
\begin{equation}
 \int_{M_2}d^2x \sqrt g R =-\int ^\infty
 _0dr\int ^{2\pi}_0d\theta\frac{A'}{A^{3/2}}\mbox{sign}(\rho')
 =\left.\frac{\pi}{\sqrt{A}}\right|^\infty
 _0=0.
\end{equation}

Therefore, we can conclude that the 2-brane-skyrmion is always
point like. This can be understood by realizing that in two
dimensions the scalar curvature integral is related to the Euler
number of the surface, which is a topological invariant and
therefore size independent, i.e., unable to give rise to a definite
size for the brane-skyrmion.

\section{Interaction Lagrangian and fermionic quantization}

In this section we will study the interaction between the
brane-skyrmions and the branons. For simplicity we will consider
only the case $M_7=M_4\times S^3$ with $\tilde
g_{\mu\nu}=\eta_{\mu\nu}$ so that $K \simeq SU(2)$. Then we can
follow the well-known steps for quantization of the standard
chiral-dynamics Skyrmion \cite{Witten1}. It is possible to split
the isometry group $G$ as $G=SU(2)_L \times SU(2)_R$ and $H$
corresponds to the isospin group $SU(2)_{L+R}$. The
parametrization of the coset is usually done in terms of an $SU(2)$
matrix $U(x)$ and the Skyrmion   is  written as
\begin{eqnarray}
U(x)&=&\exp (iF(r)\hat x^a \tau^a)=\cos F(r) + i\tau^a \hat x^a
\sin F(r)\nonumber \\ &=&\pm\sqrt{1 -\frac{\pi^2}{v^2}}+i\tau^a
\hat x^a\sin F(r),
\end{eqnarray}
where $\tau^a$ are the $SU(2)$ generators. From this expression we
can identify the Goldstone bosons fields $\pi^\alpha=v \sin F(r)
\hat x^\alpha$. The quantization of the isorotations of the
Skyrmion  solution (which correspond to rotations in the
compactified space $B=S^3$ in our case) requires the well-known
relation $J=I$ where $J$ and $I$ are the spin and the isospin
indices. In principle the allowed values of $J$ are
$J=0,1/2,1,3/2,...$. As explained by Witten \cite{Witten2},
fermionic quantization is possible because of the
Wess-Zumino-Witten (WZW) term $k \Gamma $ with $k$ integer, which 
can be added to the Goldstone boson effective action. For $k$ even
the Skyrmion is a boson, but for $k$ odd it is a fermion. For the
$SU(2)$ case, the functional $\Gamma$ has no dynamics and becomes
a topological invariant related to the homotopy group
$\pi_4(SU(2))= \bf{Z}_2$. Note that for a suitable
compactification of the space-time this is the relevant group for
the Goldstone boson map. A map belonging to the non-trivial class
could, for example, describe the creation of a Skyrmion-antiskyrmion
pair followed by a $2\pi$ rotation of the Skyrmion and finally a
Skyrmion-antiskyrmion annihilation. In the fermionic case this
field configuration must be weighted with a $-1$ in the Feynman
path integral. For an adiabatic $2\pi $ rotation of the Skyrmion
around some axis, the WZW term contributes $k \pi$ to the action
and $(-)^k$ to the amplitude which can be understood as an $exp(i
2 \pi J)$ factor. Therefore both possibilities, bosonic and
fermionic quantization of the Skyrmion, are open. In principle
this result can also be extended to the more general case of $S^N$
brane-skyrmions considered in the last section where
$M_D=M_{N+1}\times S^N$ with $D=2 N +1$ since
$\pi_{N+1}(S^N)=\bf{Z}_2$ for $N \geq 3$.

In order to study the low-energy interactions of the
brane-skyrmions with the branons, we have to obtain the appropriate
effective lagrangian. This lagrangian must be $G(B)$ symmetric and
the brane-skyrmion should be described in it by a complex field, because
of its topological charge. Thus for example this field will be a
complex scalar $\Phi$ for $J=0$ or a Dirac spinor $\Psi$ for
$J=1/2$. For the scalar case the invariant  lagrangian with the
lowest number of derivatives can be written as:
\begin{eqnarray}
{\cal L}_{s}=\alpha \Phi^* \Phi h_{\alpha\beta}(\pi)\partial_\mu
\pi^\alpha
\partial^\mu \pi^\beta.
\end{eqnarray}
The coupling $\alpha$ can be obtained from the large distance
behaviour of the branon field in the brane-skyrmion configuration.
The differential equation for $F(r)$ obtained from
the Dirac-Nambu-Goto action in the mentioned limit is the
Euler equation:
\begin{eqnarray}
r^2F''(r)+2rF'(r)-2F(r)=0.
\end{eqnarray}
The general solution of this equation is:
\begin{eqnarray}
F(r)=A r +\frac{B}{r^2}.
\end{eqnarray}
Since we are interested in those solutions in which $F(r)$
goes to zero as $r$ goes to infinity, $A$ has
to be identically zero. This  means that the general behaviour of
$F(r)$ at large distances is
\begin{equation}
F(r)\simeq \frac{B}{r^2}.
\end{equation}
Therefore in this limit:
\begin{equation}
U(x)=1+i\frac{B}{r^2}\hat x^a\tau^a+ \dots\simeq 1+i F(r)\hat
x^a\tau^a+\dots
\end{equation}
Then we have:
\begin{equation}
\pi^\alpha\simeq v \frac{B}{r^2}\hat x^\alpha,
\end{equation}
and in particular for the  Atiyah-Manton ansatz with $n_W=1$, we
get $B=L^2 \pi/2$.
 By using the
lagrangian ${\cal L}_{s}$ it is also possible to obtain the branon
field produced by the brane-skyrmion field $\Phi$ and by
comparison with the above results we arrive at \cite{Clements}
\begin{equation}
\alpha=-\frac{8}{3}\pi^2v^2B^2=-\frac{2}{3}v^2\pi^4 L_m^4.
\end{equation}
From this lagrangian it is possible, for example, to compute the
cross sections for producing a brane-skyrmion-antibrane-skyrmion
pair from two branons.

The fermionic case can be studied in a similar way \cite{Clements,Witten1}
although a consistent analysis would be more involved, since it
requires the quantization of the rotational modes. This case will be
considered elsewhere.

\section{Branon mass effects}
As shown in Section 3, if the four dimensional metric depends on
the extra coordinates, i.e. $\tilde g_{\mu\nu}(x,y)$, the branon
fields acquire a mass. In order to study the effect of the branon
masses on the brane-skyrmion, we will consider the simple case
$B=S^3$ with $\lambda=0$ and the background metric given in
(\ref{massej}). Remember that this metric corresponds to branon
fields with equal masses: $m^2=4\sigma'(0)$. In order to simplify
the calculation, we will define a new function $\tilde \sigma$ as
follows: $\tilde \sigma(m^2y^2/4)=\sigma(y^2)$. With this
definition we have $\tilde\sigma(0)=0$ and $\tilde\sigma'(0)=1$.
Using the spherical coordinates on $K$ defined in (\ref{esfk}), we
can write the background metric as:
\begin{equation}
\tilde
g_{\mu\nu}=\left(1+\tilde\sigma\left(\frac{m^2R_B^2}{4}
\sin^2(\chi_K)\right)\right)\eta_{\mu\nu},
\end{equation}
where we have used the relation (\ref{rel}).
Imposing the Skyrme ansatz in (\ref{skyan}), the induced metric on the
brane can be written as:
\begin{equation}
ds^2=g_{\mu\nu}dx^\mu dx^\nu=B(r)dt^2-A(r)dr^2-C(r)(d\theta^2
+\sin^2(\theta)d\phi^2),
\end{equation}
where:
\begin{eqnarray}
B(r)&=&1+\tilde\sigma\left(\frac{m^2R_B^2}{4}
\sin^2(F(r))\right),\nonumber \\
A(r)&=&B(r)+R_B^2(F'(r))^2,\nonumber \\ C(r)&=&r^2
B(r)+R_B^2\sin^2(F(r)).
\end{eqnarray}

Following similar steps to those in  the massless case, it is possible to
find a lower bound for the brane-skyrmion mass, thus:
\begin{eqnarray}
M_S&=&4\pi f^4\int dr\left(C(r)\sqrt{A(r)B(r)}-r^2\right)\nonumber
\\ &\geq& 4\pi f^4\int dr \left(R_B^3\sin^2(F(r))
\vert F'(r)\vert\left[1 +\tilde\sigma\left(\frac{m^2R_B^2}{4}
\sin^2(F(r))\right)\right]^{1/2}\right) \nonumber \\ &\geq& 4\pi
\vert n_W\vert  f^4 R_B^3\int_0^\pi du \sin^2 u \left[1
+\tilde\sigma\left(\frac{m^2R_B^2}{4}\sin^2 u\right)\right]^{1/2}.
\label{massm}
\end{eqnarray}

\begin{figure}[h]
\vspace*{-2cm}
\centerline{\mbox{\epsfysize=10 cm\epsfxsize=10
cm\epsfbox{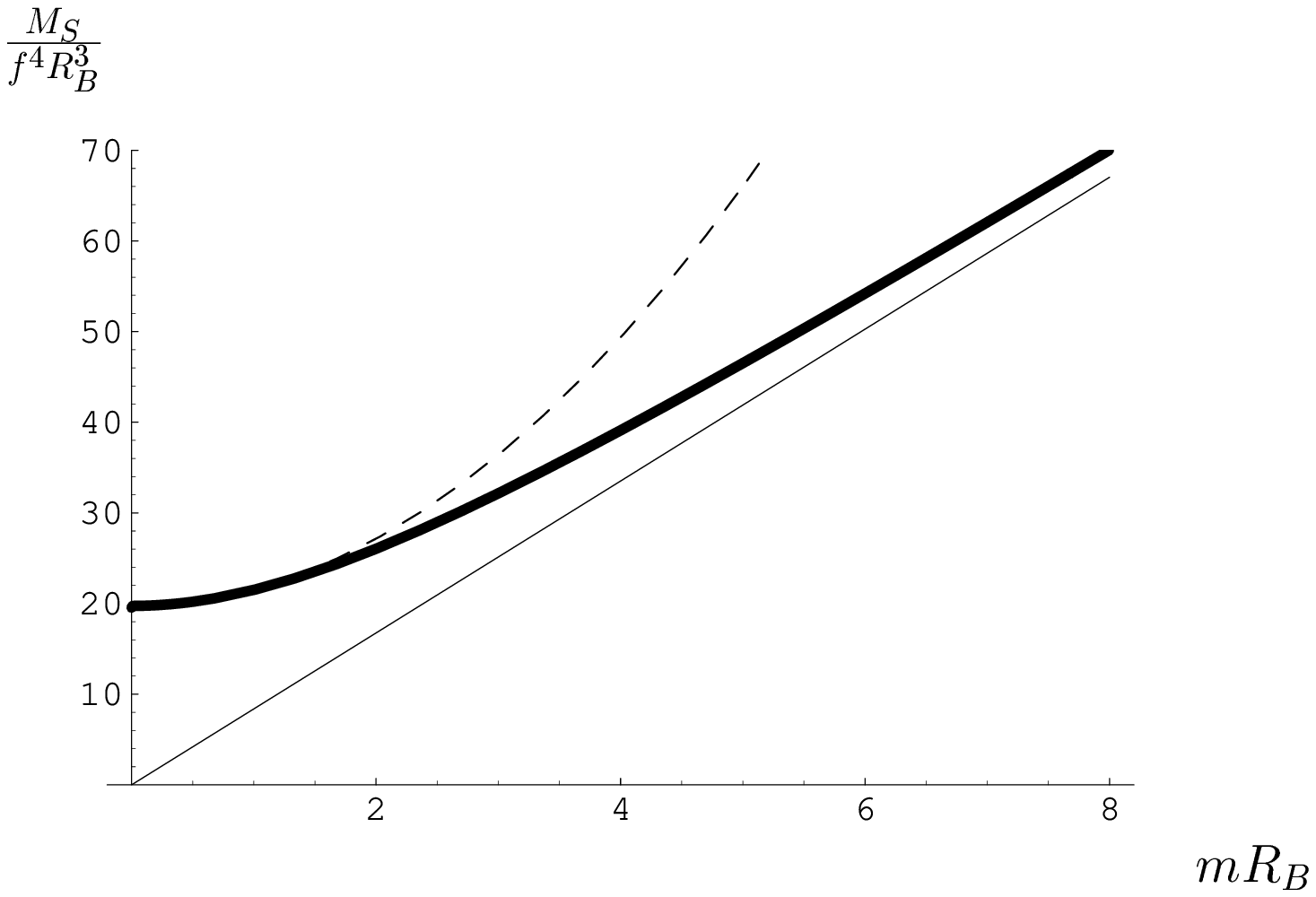}}}
%
\caption{\footnotesize{Brane-skyrmion mass $M_S$ as a function of
the branon mass $m$ in the case $\tilde\sigma(\xi)=\xi$ and
$n_W=1$. The thick continuous line represents the exact
dependence; the dashed line, the approximated behavior when $R_B
m\ll 1$; and the continuous line, the asymptotic behavior when
$R_B m\gg 1$.}}
\end{figure}

Also in the same way as in the massless case, this lower bound
coincides with the zero size brane-skyrmion mass. Therefore, the
presence of the branon masses does not affect the brane-skyrmion
stability and again we get a pointlike solution. The only
difference from  the massless case is that the brane-skyrmion mass
increases due to the $\tilde\sigma$ contribution in the last line
of (\ref{massm}).

If the branon mass is small, in  $1/R_B$ units, we  can expand
$\tilde\sigma(\xi)=\xi+\Od(\xi^2)$ in (\ref{massm}) and  obtain
that the dependence of the brane-skyrmion mass on $m$ is quadratic
at first order, for any  $\tilde\sigma$ function:
\begin{eqnarray}
M_S = 2 \vert n_W\vert \pi^2 f^4
R_B^3\left(1+\frac{3}{16}R_B^2m^2\right).
\end{eqnarray}

In the opposite limit, when the contribution of the branon masses
to the brane-skyrmion mass is more important than the topological
contribution, the brane-skyrmion mass strongly depends on  the
particular form of the  $\tilde\sigma$ function. Thus one can
find:
\begin{eqnarray}
M_S = 4\pi \vert n_W\vert  f^4 R_B^3\int_0^\pi du \sin^2 u
\left[\tilde\sigma\left(\frac{m^2R_B^2}{4}\sin^2
u\right)\right]^{1/2}.
\end{eqnarray}

Figure 6 shows these behaviours in the simple
case $\tilde\sigma(\xi)=\xi$.

\section{Wrapped states}

In this section we are going to study another kind of states which
can appear as topological excitations of the branes. These states
correspond to brane configurations  wrapped around the
compactified spaces $B$ which typically will be assumed to be
$S^N$ for $M_D=M_{N+1}\times B$. A given wrapped state is located
at some well-defined point of the space $M_N$. The possibility of
having these wrapped states is related to the homotopy group
$\pi_N(B)=\bf{Z}$ which is obviously the case for $B=S^N$, but also
for other spaces. In principle, wrapped states can be present even
when there is no world brane, i.e., when we do not have a brane
extended along the  space $M_N$. However, one of the most
interesting cases occurs when the wrapped states are located at
one point of the world brane and then can be understood as
world-brane excitations. Note that as long as the relevant homotopy
group is again $\bf{Z}$ we have also antiwrapped states which
correspond to negative winding numbers. Thus a world brane can get
excited by creating a wrapped-antiwrapped state at some given
point. In Fig.7 we show a single wrapped state at rest (left) and
branon-excited. On the left of Fig.2 we show a wrapped state
(circle) located at one point of the world brane (straight line)
for the case $N=1$.

To study in more detail the main properties of these wrapped
states we concentrate now on a four-dimensional space-time $M_4$
embedded in a $7$-dimensional bulk space that we are assuming to
be $M_7=M_4\times B$ with $M_4={\bf R}\times M_3$ and $B=S^3$.
Now, unlike the brane-skyrmion case, here the finite energy requirement
does not lead to any compactification of the world space $M_3$
because the brane is going to be wrapped around $B$ which is
compact. However, for technical reasons it is still useful to
compactify $M_3$ to $S^3$ by adding the spatial infinite point.
The wrapped brane produces the spontaneous breaking of the $M_7$
isometry group, which we assume to be $G(M_7)=G({\bf R}\times
M_3\times S^3)=G({\bf R}\times S^3)\times G(M_3)$ to the $G({\bf
R}\times S^3)\times H'$ group where $H'$ is the isotropy group of
$M_3$ which is assumed to be homogeneous. In the previous expressions,
${\bf R}$ corresponds to the time coordinate.
Thus, the coset space is
defined by $K'=G(M_3)/H'$. In the simplest case $M_3=S^3$ we have
$K'=SO(4)/SO(3)=S^3$ and then $K'\sim B\sim S^3$. Therefore the
low-energy brane excitations can be parametrized as

\begin{figure}
\vspace*{0cm}
\centerline{\mbox{\epsfysize=5.5 cm\epsfxsize=5.5
cm\epsfbox{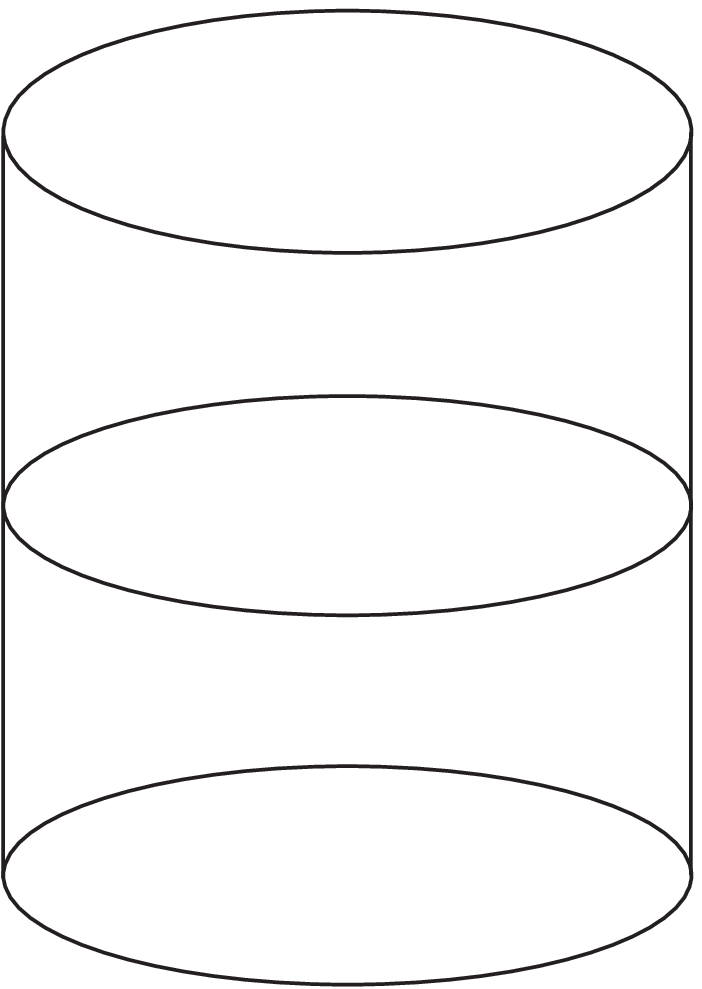}}\hspace*{2 cm}\mbox{\epsfysize=5.5
cm\epsfxsize=5.5 cm\epsfbox{cilwonp1.eps}}}
%
 \caption{ \footnotesize{Wrapped brane with topological number 1 
in $M_3=M_2\times S^1$. Its ground state  is represented 
on the left and an excited state on the right.}}
\end{figure}

\begin{eqnarray}
\Pi:B&\longrightarrow&K'\nonumber\\ y &\longrightarrow&\pi(y),
\end{eqnarray}
where  $y$ are coordinates on $B$ and $\pi$ are coordinates on
$K'$. On the other hand, as long as the quotient space $K'$ and
$M_3$ are both topologically equivalent to $S^3$, it is possible
to describe the wrapped brane by giving its position on the $M_3$
space $X^i$ as a function of $y^m$, i.e. $X^i=X^i(y^m)$. In
particular it is possible to choose the coordinates so that
\begin{equation}
X^i(y)=\frac{1}{f^2}\delta^i_\alpha \pi^\alpha(y)+...
\end{equation}
locally. In the following we will use $X^i$ instead of
$\pi^{\alpha}$ to label the wrapped-brane points in terms of the
brane parameters $y$ . Let us now rearrange the coordinates for
convenience in order to have $Y^M=(t,y^m,X^i(y))$ where $t$ is the
temporal coordinate $t=x^0$. The bulk metric is then:
\begin{eqnarray}
 G_{MN}&=&
\left(
\begin{array}{cccc}\tilde g_{00}&&0\\
&-\tilde g'_{mn}(y)&\\ 0&&-\tilde g_{ij}(x)
\end{array}\right)=\left(
\begin{array}{cccc}\tilde g'_{rs}(y)&0\\ 0&-\tilde g_{ij}(x)
\end{array}\right),
\end{eqnarray}
where $\tilde g'_{rs}$ is the background metric on the space-time
manifold ${\bf R}\times B$, i.e., $r,s=0,1,2,3$. The induced
metric on this manifold can be evaluated in a way similar to that
in  the
brane-skyrmion case. Thus in the ground state, the induced metric
on the wrapped brane is given by the four-dimensional components
of the bulk space metric, i.e. $g'_{rs}=\tilde g'_{rs}=G_{rs}$.
When branons are present, the induced metric is given by
\begin{equation}
g'_{rs}=\tilde g'_{rs}-\partial_{r}X^i\partial_{s}X^j\tilde g_{ij}
\end{equation}
and the square root of the induced metric determinant can be
written as
\begin{equation}
\sqrt{g'}=\sqrt{\tilde g'}\left(1-\frac{1}{2}\tilde g'^{rs}\tilde
g_{ij}\partial_rX^i\partial_sX^j+...\right).
\end{equation}
On the other hand, the action including the scalar curvature term is
given by:
\begin{equation}
S_B=-f^4 \int_{{\bf R}\times B}dtd^3y\sqrt{g'}+\lambda f^2
\int_{{\bf R}\times B}dtd^3y\sqrt{g'}R',\label{wra}
\end{equation}
where $R'$ is the induced curvature on the wrapped brane
and the volume term is now finite for fixed time. For
small excitations, the effective action becomes
\begin{equation}
S_{eff}[X]=S_{eff}^{(0)}[X]+ S_{eff}^{(2)}[X]+ ...
\end{equation}
where the effective action for the branons up to $\Od(p^2)$ is
nothing but the non-linear sigma model corresponding to a
symmetry-breaking pattern $G(M_3)\rightarrow H'$:
\begin{eqnarray}
S_{eff}^{(0)}[X]&=&-f^4  \int_{{\bf R}\times B}dtd^3y\sqrt{\tilde
g'},
\end{eqnarray}
\begin{eqnarray}
 S_{eff}^{(2)}[X]&=&\frac{f^4}{2} \int_{{\bf R}\times B}dtd^3y
\sqrt{\tilde g'}\tilde
 g_{ij}\tilde
 g'^{rs}\partial_{r}X^i\partial_{s}X^j
+\lambda f^2 \int_{{\bf R}\times B}dtd^3y\sqrt{\tilde g'} \tilde
R',\hspace{1 cm}
\end{eqnarray}
where $\tilde R'$ is the background curvature on the wrapped
brane, without excitations. Notice that $i, j,...$  are $M_3$
indices, whereas $r, s,...$ are indices on the ${\bf R}\times K'$
manifold. This effective action is again an expansion in powers of
$p \sim \partial_r X\sim\partial_r g'/f$, i.e. it is a low-energy
expansion. For static configurations the mass, to the lowest
order, is given from (\ref{wra}) again by
\begin{equation}
M_W=f^4 \int_{B}d^3y\sqrt{g'}.
\end{equation}
The minimum is found for $X^i=0$:
\begin{eqnarray}
M_W=2\pi^2f^4R_B^3,
\end{eqnarray}
which is proportional to the $B=S^3$ volume as expected. In this
case we have the brane wrapped around $B$ with the minimal
possible brane volume. For small enough $\lambda$,
adding the curvature term does not change the picture very much:
\begin{equation}
M_W=f^4 \int_{B}d^3y\sqrt{g'}-\lambda f^2 \int_{B}d^3y\sqrt{g'}R'.
\end{equation}
Since the scalar curvature on a 3-sphere is $\tilde
R'=-6/R_B^2$, we find
\begin{eqnarray}
M_W=2\pi^2f^4R_B^3\left(1+6\frac{\lambda}{R_B^2f^2}\right).
\end{eqnarray}
Thus the brane is still wrapped minimizing its volume, but we have
a new contribution to the mass coming from the brane curvature
which coincides with the $B$ curvature. This result obviously
applies to branes wrapped once around $B$. The generalization to the 
cases where the brane is wrapped $n_W\in \bf{Z}$ times is
straightforward  resulting just in a factor of $\vert n_W \vert$
in the above equation.

It is very interesting to realize that the value obtained  for
the wrapped-state mass is exactly the same previously given for
the brane-skyrmion mass in Table.1, as the upper bound for
positive $\lambda$ and the exact value for negative $\lambda$,
provided $\lambda > -R_B^2 f^2/6$. The fact that our brane
action is  defined in an entirely geometrical way, makes it possible
to give a beautiful explanation of this fact. In order to have a
graphical picture of this explanation it is useful to consider for
a moment the $N=1$ case. As we have already noticed, the
brane-skyrmion is not stable in this case, but still we can ignore
this fact and use the $N=1$ geometry as an abstract representation
of the $N=3$ case. Notice however that wrapped states are stable
even for $N=1$.

On the right of  Fig.2 we have represented a brane-skyrmion
corresponding to a positive value of $\lambda$. According to our
previous discussion the brane-skyrmion has a non-zero size. This
makes possible to pass through the brane from one side to the
other, showing the  topological defect as some kind of hole in the
brane. The mass of the brane-skyrmion has volume and  curvature
contributions. As long as both of them are positive, the curvature
term avoids the generation of the singularities present in
zero-size Skyrmions. When $\lambda$ goes to
zero, the curvature contribution vanishes and the brane-skyrmion
collapses to zero size. The brane configuration is then
represented in the left side of Fig.2. It is also interesting that
this picture could also represent a wrapped state (circle) plus a
world brane (straight line). Then we realize that the shape, size
and curvature are exactly the same for both configurations and this
is the reason why the mass of the brane-skyrmion  equals the
wrapped state mass in this case (notice that the brane-skyrmion
mass was defined as the corresponding brane-skyrmion configuration
mass minus the brane-world mass in order to have a finite value)
i.e., it is proportional to the $B$ volume. In spite of this, the two 
configurations are not the same because  their topology is
different. The brane-skyrmion  is extended on  both the
compactified $M_3$ space and the extra-dimensional space $B$, but
the wrapped states  only around the extra dimension space $B$.
Thus, brane-skyrmions are classified according to the homotopy
classes of the mappings $\Pi: M_3 \rightarrow K$, whereas wrapped
states are labeled by the number of times the brane wraps around
the extra dimensions. Another way to understand why they are
different is to realize that the brane-skyrmion is made of a
single piece, unlike the wrapped configuration which has two
different pieces (the wrapped brane and the world brane). Thus
they cannot be connected by a classical process, although quantum
tunneling could produce in principle transitions between one to
the other. For small and negative $\lambda$, the volume term
still dominates but the curvature term produces a negative
contribution to the mass. Both terms are proportional to the
volume but the second one is also proportional to the curvature
and $\lambda$. This result clearly applies to the brane-skyrmion
and wrapped brane-states simultaneously. Finally, for $\lambda <
-R_B^2 f^2/6$ the curvature term dominates and it is energetically
favored for them to wrap, making the mass functional unbounded
from below.

Now we can consider excitations (branons) of the wrapped ground
state. For small excitations the relevant action is
\begin{eqnarray}
 S_{eff}[X]=\frac{f^4}{2} \int_{{\bf R}\times B}
 dtd^3y\sqrt{\tilde g'}\sum_{i=1}^3\{\partial_t X^i\partial_t X^i-\tilde
 g'^{mn}\partial_{m}X^i\partial_{n}X^i\},
\end{eqnarray}
which describes three free scalars $X^i$
propagating on an $S^3$ manifold. The corresponding spectrum is
well known \cite{Birrel} and we have for each field:
\begin{eqnarray}
E_{n_i}=\frac{1}{R_B}\sqrt{n_i^2-1}.
\end{eqnarray}
The different states are labeled by $(n_i,j_i,m_i)$; $n_i=1,2,3,4,...$
; $j_i=0,1,2,...,n_i-1$; $m_i=-j_i,-j_i+1,...,j_i$. For a given $n_i$
the degeneracy for each free scalar is
\begin{eqnarray}
g_{n_i}=\sum_{j_i=0}^{n_i-1}(2j_i+1)=n_i^2 .
\end{eqnarray}
On the other hand the different topological sectors are labeled by
$n_W=0,\pm 1,\pm 2, \pm 3,...$ and the corresponding masses (for
moderate negative $\lambda$) are
\begin{eqnarray}
M_W=|n_W|2\pi^2f^4R_B^3\left(1+6\frac{\lambda}{R_B^2f^2}\right),
\end{eqnarray}
so that the degeneracy in this case is $g_{n_W}=2$
 corresponding
to the two different orientations.

All the above discussion can be extended without any difficulty to
the general case $M_D={\bf R}\times M_N \times S^N$ with $D=2 N+
1$ where we can have N-brane wrapped states. In this case we will
have that the small oscillations over the ground state can be
described as $N$ free scalars propagating on a $S^N$ manifold.
Then the energy spectrum for each field is
\begin{eqnarray}
E_{n_i}=\frac{1}{R_B}\sqrt{(n_i-1)(n_i-2+N)},
\end{eqnarray}
where $n_i=1,2,3,4,... ;i=1,2,3,... N$. In this case, the
degeneracy for each free scalar is
\begin{eqnarray}
g_{n_i}=\frac{(2n_i+N-3)(n_i+N-3)!}{(N-1)!(n_i-1)!}.
\end{eqnarray}
The energy of the winding modes ($\pi_N(S^N)=\bf{Z}$) for small
negative curvature parameter
\begin{eqnarray}
M_W=|n_W|\frac{2\pi^\frac{N+1}{2}}{\Gamma(\frac{N+1}{2})}f^{N+1}R_B^N
\left(1+N(N-1)
\frac{\lambda}{R_B^2f^2}\right)
\end{eqnarray}
and again $g_{n_W}=2$.

In addition to these states it is well known that, due to the compact nature of B,
we always have the standard Kaluza-Klein
spectrum for the particles or topologically trivial branes (in the
sense of $B$)  propagating along the compactified dimensions. For example, 
for $N=1$, we have, in addition to the 1-branes (strings) wrapped on
$B=S^1$, the corresponding Kaluza-Klein spectrum which is given by 
\begin{eqnarray}
M_{KK}=\frac{\vert n'\vert }{R_B},
\end{eqnarray}
where $n'\in \bf{Z}$ and $g_{n'}=2$ except for $g_{0}=1$. For the
winding states ($\pi_1(S^1)=\bf{Z}$) we have
\begin{eqnarray}
M_W=|n_W|2\pi R_B f^2.
\end{eqnarray}
Thus we recover the well-known string T-duality (exchange of
Kaluza-Klein and winding modes) by making the replacements
\begin{eqnarray}
2\pi
R_Bf^2&\longleftrightarrow&\frac{1}{R_B}\nonumber\\n_W&\longleftrightarrow&n' .
\end{eqnarray}
Obviously for higher $N$ this duality is not expected to apply,
since the degeneracy of the different kind of states
(Kaluza-Klein and topological) does not fit.

\section{Summary and conclusions}

In the brane-world scenario with $f\ll M_D$ (brane tension much
smaller than the fundamental scale of D-dimensional gravity), the
relevant low-energy excitations of the brane correspond to the
Goldstone bosons (branons) associated with the spontaneous breaking
of the compactified extra-dimensions isometries
produced by the brane.

Assuming a very general form for the brane action (volume plus
curvature term), it is possible to derive in a systematic way the
low-energy effective lagrangian for the branons which has the
typical form of a non-linear sigma model with well-defined arbitrary
 higher-derivative terms.

Under suitable assumptions about the third homotopy group of the
space $B$, this effective action gives rise to a new kind of states
corresponding to topological defects of the brane
(brane-skyrmions) which are stable whenever the curvature
parameter $\lambda$ is not too negative. The mass and the size of
the brane-skyrmions can be computed in terms of the brane tension
scale ($f$), $\lambda$ and the size of the space $B$ ($R_B$). The
brane-skyrmions can be understood as some kind of holes in the
brane that make it possible to pass through them along the $B$ space.
This is because in the core of the topological defect the symmetry
is restablished. In the case considered here the broken symmetry
is basically the translational symmetry along the
extra-dimensions. Thus the core of the brane-skyrmion plays the
role of a window through the brane, which is a nice geometrical
interpretation of this object. For $\lambda =0$ or negative the
brane-skyrmion collapses to zero size
and that window is closed.

Brane-skyrmions can in principle be quantized as bosons or
fermions by adding a Wess-Zumino-Witten-like term to the branon
effective action. This is a very interesting possibility since it
provides a completely new way of introducing fermions on the
brane. The low-energy effective lagrangian describing the interactions
between branons and brane-skyrmions can also be obtained in a systematic
way. This opens the door for the study of the possible
phenomenology of these states at the Large Hadron Collider (LHC)
currently under construction at CERN.

The effects on the brane-skyrmions of a possible small branon mass
due to a explicit breaking of the translational invariance by the
bulk metric, have also been considered.

We have studied another different set of states corresponding
to a brane wrapped on the extra-dimension space $B$ (wrapped
states) and we have analysed their connection with the
brane-skyrmion states.

Finally, we have extended also our study to the case of higher
dimensions where similar results hold. We understand that this
could have some relevance in the context of pure M-theory where
solitonic 5-branes are present which could wrap around
5-dimensional spheres.

We understand that the brane-skyrmions and wrapped states studied
in this paper are quite interesting objects (both from the
theoretical and perhaps from a more phenomenological point of
view) and thus we think that they deserve further research. Work
is in progress in this direction.

\section*{Appendix}
For small brane excitations in a background metric $\tilde
g_{\mu\nu}$, the effective action (\ref{Nambu4}) can be expanded
in branon fields derivatives as follows:
\begin{equation}
S_{eff}[\pi]=S_{eff}^{(0)}[\pi]+ S_{eff}^{(2)}[\pi]+
S_{eff}^{(4)}[\pi]+ ...
\end{equation}
where:
\begin{eqnarray}
S_{eff}^{(0)}[\pi]&=&-f^4 \int_{M_4}d^4x\sqrt{\tilde g}.
\end{eqnarray}
The $\Od(p^2)$ contribution is the non-linear sigma model
corresponding to a symmetry-breaking pattern $G\rightarrow H$ plus
the background scalar curvature term:
\begin{eqnarray}
 S_{eff}^{(2)}[\pi]&=&\frac{1}{2}\int_{M_4}d^4x\sqrt{\tilde g}
h_{\alpha\beta}\partial_{\mu}\pi^\alpha\partial^{\mu}\pi^\beta
+\lambda f^2\int_{M_4}d^4x\sqrt{\tilde g} \tilde R.
\end{eqnarray}
We are assuming that the branons derivative terms are of the same
order as those with metric derivatives. The fourth-order term is
obtained by expanding both the metric determinant and the induced
scalar curvature in branon fields:
\begin{eqnarray}
 S_{eff}^{(4)}[\pi]&=&\frac{-1}{8f^4}\int_{M_4}d^4x\sqrt{\tilde g}h_{\alpha\beta}
 h_{\gamma\delta}(\partial_{\mu}\pi^\alpha\partial^{\mu}\pi^\beta\partial_{\nu}
 \pi^\gamma\partial^{\nu}\pi^\delta-2\partial_{\mu}\pi^\alpha\partial^{\nu}
 \pi^\beta\partial_{\nu}\pi^\gamma\partial^{\mu}\pi^\delta)\nonumber\\
&+&\frac{\lambda}{2f^2} \int_{M_4}d^4x\sqrt{\tilde
g}h_{\alpha\beta}\partial^{\mu}\pi^\alpha\partial^{\nu}\pi^\beta(2\tilde
R_{\mu\nu}-\tilde R\tilde g_{\mu\nu})\nonumber\\
&+&\frac{\lambda}{f^2}\int_{M_4}d^4x\sqrt{\tilde
g}\Delta_\mu\{\partial^{\nu}\pi^\alpha\Delta_\xi\partial^{\eta}\pi^\beta
h_{\alpha\beta}(\tilde g^{\xi\nu}\tilde g^{\mu\eta}-\tilde
g^{\xi\eta}\tilde g^{\nu\mu}) \}, \label{ese4}
\end{eqnarray}
where
\begin{eqnarray}
\Delta_\mu\; T^{\alpha\beta \;...\;\nu\rho ...}_{\epsilon\delta\;
... \;\sigma\tau \; ...} \equiv \tilde D_\mu \;T^{\alpha\beta
\;...\;\nu\rho ...}_{\epsilon\delta\; ... \;\sigma\tau \; ...}+
\partial_{\mu}\pi^\gamma \hat D_{\gamma}T^{\alpha\beta \;...\;\nu\rho
...}_{\epsilon\delta\; ... \;\sigma\tau \; ...}
\end{eqnarray}
and $T^{\alpha\beta \;...\;\nu\rho ...}_{\epsilon\delta\; ...
\;\sigma\tau \; ...}$ is an arbitrary tensor with indices in both
spaces $M_4$ and $K$. Here $\tilde D_\rho$ is the covariant
derivative in $M_4$ with Christoffel symbols
($\tilde\Gamma^\nu_{\rho\mu}$) corresponding to $\tilde
g_{\mu\nu}$, and $\hat D_\gamma$ refers to the covariant
derivative in $K$ with Christoffel symbols
($\hat\Gamma^\alpha_{\gamma\beta}$) defined from
$h_{\alpha\beta}(\pi)$. Thus, for example:
\begin{eqnarray}
\Delta_\rho (\partial_{\mu}\pi^\alpha)=
\partial_\rho(\partial_{\mu}\pi^\alpha)-
\tilde\Gamma^\nu_{\rho\mu}\partial_{\nu}\pi^\alpha
+\hat\Gamma^\alpha_{\gamma\beta}
\partial_{\rho}\pi^\gamma\partial_{\mu}\pi^\beta.
\end{eqnarray}

Notice that $\mu, \nu,...$  are $M_4$ indices, whereas
$\alpha,\beta,...$ are indices on the $K$ manifold. Let us
emphasize again that the above effective action is an expansion in
branon fields (or metric) derivatives over $f^2$ and not an
expansion in powers of $\pi$ fields, i.e, it is a low-energy
effective action. The last term in (\ref{ese4}) is a total
divergence, and therefore it does not contribute to the branon
equations of motion.

\vspace{.5cm}
 {\bf Acknowledgements:} This work was 
partially supported by the Ministerio de Educaci\'on y
Ciencia (Spain) (CICYT AEN 97-1693 and PB98-0782).    \\

\end{document}